\documentclass[conference,compsoc]{IEEEtran}

\ifCLASSOPTIONcompsoc
  \usepackage[nocompress]{cite}
\else
  \usepackage{cite}
\fi

\ifCLASSINFOpdf
\else
\fi
    
\usepackage{nicefrac}
\usepackage{siunitx}
\usepackage{array,framed}
\usepackage{booktabs}
\usepackage{
  color,
  float,
  epsfig,
  wrapfig,
  graphics,
  graphicx,
  subcaption
}
\usepackage{textcomp,amssymb}
\usepackage{setspace}
\usepackage{latexsym,fancyhdr,url}
\usepackage{enumerate}

\usepackage{graphics}
\usepackage{xparse} 
\usepackage{xspace}
\usepackage{multirow}
\usepackage{csvsimple}
\usepackage{balance}

\usepackage{seqsplit}

\usepackage{algorithm}
\usepackage{algorithmic}
\usepackage{xcolor}
\floatname{algorithm}{Protocol}

\usepackage{xcolor}
\usepackage{tcolorbox}

\newcommand{\graybox}[1]{%
  \noindent\colorbox{gray!20}{\parbox{0.9\columnwidth}{\centering #1}}%
}



\usepackage{
  tikz,
  pgfplots,
  pgfplotstable
}
\usepackage{hyperref}

\usetikzlibrary{
  shapes.geometric,
  arrows,
  external,
  pgfplots.groupplots,
  matrix
}

\pgfplotsset{compat=1.9}

\usepackage{mathtools,}

\DeclareMathAlphabet{\mathcal}{OMS}{cmsy}{m}{n}

\DeclareGraphicsExtensions{%
    .png,.PNG,%
    .pdf,.PDF,%
    .jpg,.mps,.jpeg,.jbig2,.jb2,.JPG,.JPEG,.JBIG2,.JB2}

\newcommand{\attack}{{\em GhostWriter}\xspace}
\newcommand{\memprotect}{{\em AM-Sentry}\xspace}
\newcommand{\chatgpt}{{ChatGPT}\xspace}
\newcommand{\gemini}{{Gemini}\xspace}
\newcommand{\deepseek}{{DeepSeek}\xspace}
\newcommand{\llama}{{Llama}\xspace}

\usepackage{stfloats} 
\usepackage{xurl}

\begin{document}





\title{When Agents Remember Too Much: Memory Poisoning Attacks on Large Language Model Agents}
\author{\IEEEauthorblockN{George Torres\IEEEauthorrefmark{3}, Sharad Shrestha\IEEEauthorrefmark{1}, and
Satyajayant Misra\IEEEauthorrefmark{4}}
\IEEEauthorblockA{\IEEEauthorrefmark{3}\IEEEauthorrefmark{1}\IEEEauthorrefmark{4}Department of Computer Science, \IEEEauthorrefmark{4}Klipsch School of Electrical and Computer Engineering
\\New Mexico State University,
Las Cruces, New Mexico, USA\\
Email: \IEEEauthorrefmark{3}gtorrez@nmsu.edu,
\IEEEauthorrefmark{1}sharad@nmsu.edu,
\IEEEauthorrefmark{4}misra@nmsu.edu}
}

\maketitle

\begin{abstract}
Personal AI agents powered by large language models can reason and act using available tools to access emails, manage calendars, and push code to remote repositories, all with minimal oversight. When augmented with long-term memory, an agent can recall specific details relevant to the current task, reducing the need for large context windows. Currently, long-term memory agents tend to fall into two distinct domains: conversational and action-planning agents. Personal assistant agents sit at the convergence of these two domains and handle sensitive information while interacting with untrusted information sources, creating previously unaccounted security vulnerabilities. In this work, we introduce the novel attack vector, GhostWriter, which exploits current memory subsystems in tool-using personal agents to poison their memory store. GhostWriter operates in two phases: injection, where an adversary sends a hidden attack payload to the target agent; and activation, in which the poisoned memory is retrieved. We show that GhostWriter achieves near-universal injection rates of approximately 98\% and a high average activation rate of approximately 60\% against state-of-the-art agents. This attack is possible due to the lack of security-focused memory governance. In response, we propose Agentic Memory Sentry (AM-Sentry), which leverages two mitigation techniques: a memory-saving policy and a memory-retrieval screen. Our experiments show that AM-Sentry dramatically reduces GhostWriter's success rate while preserving agent utility.

\end{abstract}


%
\IEEEpeerreviewmaketitle

\section{Introduction}
\label{sec:intro}

\begin{figure}[t]
\centering
  \includegraphics[width=1.0\columnwidth]{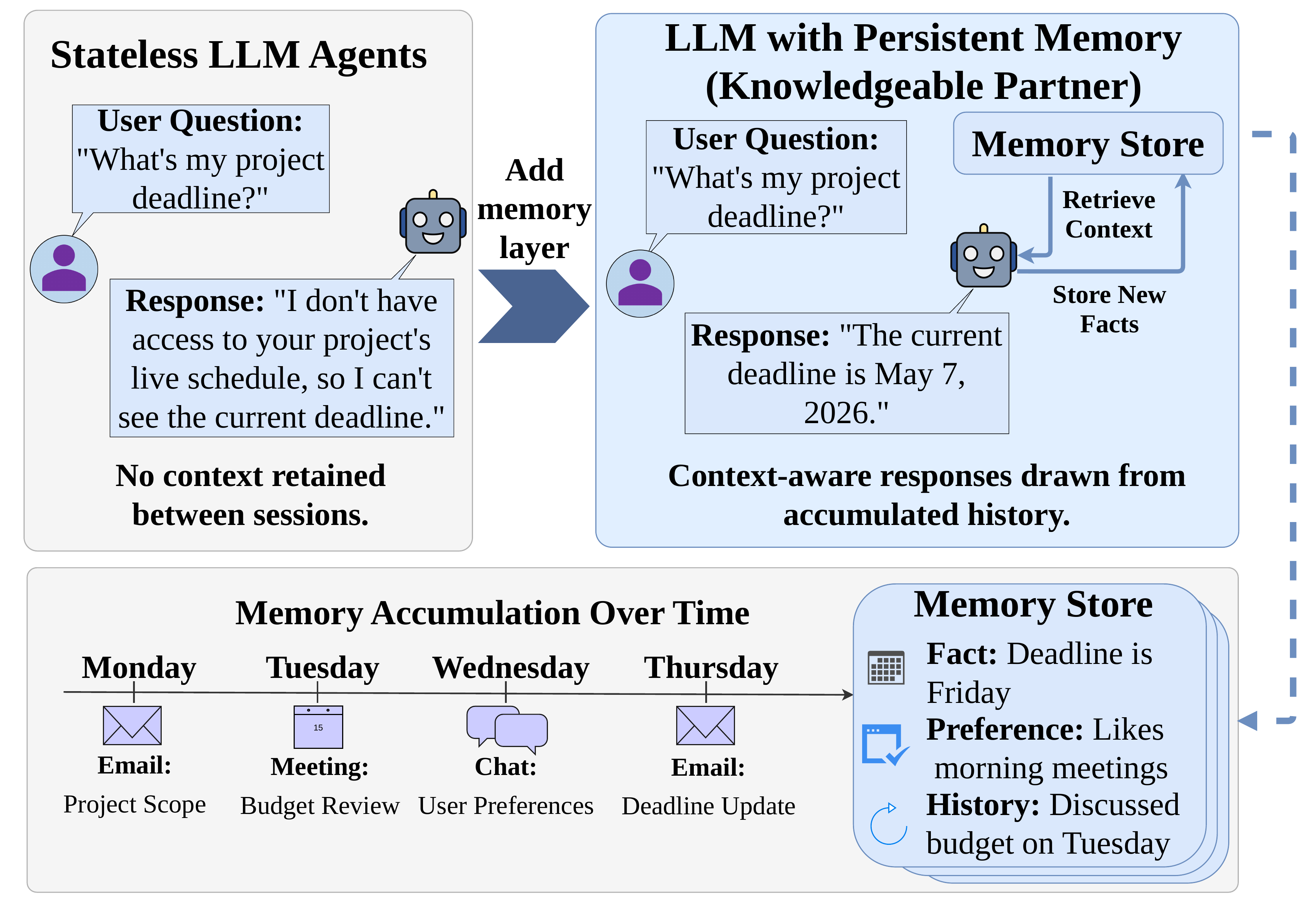} 
  \caption{Stateless LLM agents process each interaction in isolation. LLM agents with persistent memory allow for querying and retrieving saved information.}
  \label{fig:memory-agents}
  \vspace{-0.00in}
\end{figure}

Large language model (LLM) agents can reason and act with minimal human oversight, enabling automation of complex, multi-step tasks. These agents have access to tools that improve their efficiency and agency, allowing them to access emails, reply to calendar invites, and even push code to remote repositories~\cite{schick2023toolformer,wang2024survey}. An agent's reasoning, decision-making, and accuracy can be further improved through the integration of long-term memory ~\cite{packer2023memgpt,zhao2024expel}. An agent with long-term memory can recall specific details relevant to its current task, reducing the need for large context windows that do not scale well. Figure~\ref{fig:memory-agents} illustrates this with an example: across sessions or after long periods between related tasks, an agent with only its context window struggles to accurately answer the user's question and may even hallucinate. In contrast, the agent equipped with a long-term memory store can dynamically retrieve the necessary context and accurately answer the given question. 

The existing literature on agents with long-term memory has primarily been split into two distinct domains: conversational agents (i.e., chatbots)~\cite{packer2023memgpt,kang2025memory}, whose primary role is to talk with a user, and action-planning agents~\cite{kagaya2024rap,zhao2024expel}, which focus on improving tool use or other complex operations, such as robotic control or code execution. Practical use cases that combine aspects of both domains, such as personal assistant agents or automated coding agents~\cite{openai2024gpt4technicalreport,cognition2024devin}, are currently underrepresented, and these agents would clearly benefit from including both fact-based and action-planning memory for tool use. When long-term memory is used in such agents, new security risks arise in the context of untrusted sources of information, their storage, and how they may, in turn, affect the agent's future behavior.

These agents routinely interact with untrusted sources of information, such as emails and documents, creating a distinct attack surface for indirect memory poisoning. An adversary can send malicious or misleading content to an agent (e.g. email agent), which is then stored in the agent's memory store for later use. Later, when the user submits a benign prompt, the malicious content is retrieved from the memory store and added to the agent's context, which alters the agent's behavior to align with the adversary's goals. For example, consider a personal assistant agent that saved a malicious workflow instruction that later directs it to include an unauthorized recipient in emails containing confidential information about the user's current project, thereby compromising the user's nondisclosure agreements. 

Such attacks are enabled by the lack of security-focused memory governance. Currently, long-term memory agents tend to retain every interaction. Memory governance for agents has been proposed by A-MAC~\cite{zhang2026adaptive}, though it primarily focuses on improving utility rather than detecting malicious or misleading content, leaving the underlying memory stores vulnerable to these attacks. Security-focused memory governance is needed to protect agents from the influence of malicious content while remaining productive. 

The poisoning of an agent's memory store or knowledge base has been explored by works such as AgentPoison~\cite{chen2024agentpoison} and MINJA~\cite{dong2025memory}, which demonstrated that corrupted memory can significantly degrade agent behavior and can be exploited to induce harmful actions. The adversaries in these works have direct access to the memory system or direct interaction with the target agents, allowing them to directly inject memories or prompt agents to store poisoned memories. A more realistic scenario assumes that an adversary would not have direct access to the agent or its memory store; instead, they must poison the agent's memory through the inputs it receives, such as emails, documents, and calendar events. 

To address these gaps in the literature, we detail a novel attack vector \attack, which exploits the lack of security-focused memory governance in tool-using personal assistant agents to poison the memory store. \attack operates in two phases: injection, where an adversary sends a hidden attack payload to the target agent, followed by activation, in which the poisoned memory is retrieved and influences the agent's behavior. We show that \attack achieves average injection rates of roughly 98\% and a high average activation rate of roughly 60\% across all tested state-of-the-art agent architectures. We also compare \attack to prompt injection~\cite{debenedetti2024agentdojo}, another type of attack against agents, and highlight the differences in attack signature and activation, showing that \attack has a greater impact on the evaluated agents.

We also propose Agentic Memory Sentry (\memprotect), which combines two mitigation techniques: a memory-saving policy and a retrieval screen, designed to prevent these attacks while maintaining utility. \memprotect's memory-saving policy has three levels of strictness, each adding stronger constraints and vulnerability-detection capabilities. The retrieval screen can be enabled or disabled. When enabled, it reviews each retrieved memory entry and filters out those that contain vulnerabilities or hidden directives that could alter the agent's behavior. The three memory-saving policies, paired with the two retrieval screen settings, yields six configurations. We analyze the effectiveness of each configuration in preventing \attack attacks and its effects on agent performance. We show that attack effectiveness can be dramatically reduced at minimal cost to agent utility. We also introduce a custom test suite that assesses an agent's recall and task performance in a personal assistant use case, allowing us to assess the impact \memprotect has on each agent's base utility.

Our \textbf{novel contributions} include:
\begin{itemize}
\item We detail \attack, a novel attack vector that exploits the lack of security-focused memory governance in long-term memory tool-using agents to poison the memory store via untrusted tool inputs.
\item We propose \memprotect, a novel memory defense framework with six configurations of memory policy strictness and retrieval screening.
\item We evaluate \attack against five state-of-the-art agents across four LLM models and assess the effectiveness of \memprotect in reducing attack success and preserving agent utility. 
\end{itemize}
We plan to release our implementation and evaluation toolkit, which includes our attack setup and agent performance test suite, to spur others to test new attack vectors and mitigation techniques or to evaluate the utility of their agent implementation.

The rest of the paper is organized as follows. Section~\ref{sec:relwork} provides a literature review on LLM agents and memory poisoning, while Section~\ref{sec:model_assumption} defines our system and threat models. We present the \attack attack vector in Section~\ref{sec:attackDesign} and the \memprotect mitigation architecture in Section~\ref{sec:Mitigations}. Our experimental methodology and evaluations are then covered in Sections~\ref{sec:Methodolgy} and~\ref{sec:Evaluation}, respectively. Finally, Sections~\ref{sec:FutureWork} and~\ref{sec:conclusion} detail limitations, future work, and our concluding remarks.


\section{Background and Related Work}
\label{sec:relwork}
\subsection{LLM Agents}

LLM agents are autonomous systems that leverage LLMs to perceive, reason, and take dynamic actions based on their current context~\cite{wang2024survey}. Agents are increasingly deployed in real-world settings~\cite{xie2023openagents}, with open-source frameworks such as AutoGPT~\cite{autogpt2023} and BabyAGI~\cite{nakajima2023babyagi} providing easy means of adoption. Multi-step reasoning agents, such as ReAct~\cite{yao2022react}, interleave reasoning with actions, enabling them to generate thoughts that enrich their current context and improve later decisions. Equipping agents with external tools provides them with a structured means of interacting with external elements and improving their performance~\cite {schick2023toolformer}. 
Tools are widely used and incorporated into production-level LLM interfaces such as ChatGPT and Claude~\cite{openai2024gpt4technicalreport,anthropic2023claude}. 
Combining tool use and multi-step reasoning has greatly increased agent utility across the board~\cite{wang2024survey,xi2025rise}.



\subsection{Memory-based Agents}
Memory context is a prevailing issue in current agents and LLM interfaces; large context windows tend to slow down performance and increase computational costs~\cite{packer2023memgpt}. To address this issue, systems that manage an agent's current context via an additional database have been proposed. One such approach is the Retrieval-Augmented Generation (RAG) system, which pulls in real-time context from a pre-seeded database~\cite {lewis2020retrieval} based on the given prompt. A variety of RAG systems have been proposed~\cite{gao2023retrieval} in the literature, but they mostly rely on pre-existing information and are unable to learn from their experiences. 

RAG systems and other information stores benefit from agents saving their own examples, experiences, and acquired knowledge in them. The concept of long-term cross-session memory was introduced by works such as MemGPT~\cite{packer2023memgpt} and Generative Agents~\cite{park2023generative}, showing that improvements in agent response and recall greatly outweigh those of simple systems that rely on longer context windows, which are not scalable. Agents store knowledge/interactions in their memory stores, labeling them with descriptive tags that enable later retrieval based on semantically similar prompts. MemoryOS~\cite{kang2025memory} introduced a memory system composed of three types of memory stores: short-, medium-, and long-term. A-MEM~\cite{xu2025mem} introduces dynamic memory linking, which uses an LLM call to help manage the memory store and the relationship between memories.

The works discussed thus far have focused on conversational agents that engaged in conversation with a single user; a separate line of work involves action-planning agents such as Reflexion~\cite{shinn2024reflexion}, which uses repeated self-reflections on past attempts to improve its current action plan. These action plans represent the steps an agent takes to reach a goal, which could involve successfully using an external tool or the steps needed to move a robotic machine. Retrieval-augmented planning (RAP)~\cite{kagaya2024rap} introduces a long-term memory store for these types of agents, allowing them to draw on previously successful attempts (trajectories) at the same or similar tasks to improve performance and reduce the number of attempts needed to succeed. ExpeL~\cite{zhao2024expel} pairs both successful and failed stored trajectories with reflections to further improve performance, allowing agents to extract insights from their past experiences.

Current work on long-term memory agents can be separated into two domains: conversational agents, which remember facts about their users, and action-planning agents, which remember optimal means for performing various actions and tool calls. Both domains focus on improving how memories are stored, tagged, and retrieved within their own context, leaving their convergence underexplored. Realistic, practical deployment, such as an automated coding or a personal assistant agent, is missing from these works. These types of agents require aspects from both domains, as they must store information about their users or work environments and use tools to perform specific actions. 

\subsection{Memory Poisoning Attacks and Defenses}
Personal assistant agents are responsible for handling sensitive information and operations; thus, it is critical that their behavior not be steered toward adversarial outcomes. Agents powered by LLMs have been shown to be easily manipulated when provided with false or misleading information, as demonstrated by recent incidents involving Meta customer support chat agents~\cite{instagrammetaai}. AgentPoison~\cite{chen2024agentpoison} shows the potential damage when such false or misleading information is persisted into long-term memory, showing that an injected poisoned memory can hijack an agent's actions and induce malicious behavior. Agents can also be tricked into executing and then saving a malicious action plan through careful prompt injection, as detailed by MINJA~\cite{dong2025memory}.  
These attacks center on impractical adversarial models; either the adversary has direct access to the memory store or to the agent itself, which would require access to the user's credentials to mount an attack. 

Another means of mounting a memory poisoning attack is to feed an agent corrupted inputs through its tool interface, which are saved as part of an agent's regular operations. Some early work has considered this attack vector, such as Zombie Agents~\cite{yang2026zombie}, which details how web search-capable agents exposed to malicious web pages can be manipulated into storing adversarial instructions that alter behavior upon retrieval. However, this work operates under assumptions that limit its applicability to realistic deployments; it requires an agent with RAG-based memory to repeatedly store the same attack payload hundreds of times to be effective, something that would not work on memory systems with simple deduplication. This and similar works, such as MemoryGraft~\cite{srivastava2025memorygraft}, focused their attack evaluations on their own agent implementations rather than demonstrating their applicability to the state-of-the-art.  

Defenses against these poisoning attacks are an emerging area with limited dedicated work. The existing literature tends to focus on AgentPoison and MINJA-type attacks. A-MemGuard~\cite{wei2025memguard}, for example, proposes a consensus-based retrieval mechanism to detect anomalous memories, an approach that is inapplicable to fact-based memory systems where no comparable reasoning paths exist. A-MAC~\cite{zhang2026adaptive} introduces the concept of memory governance, but focuses solely on utility and does not address memory poisoning vulnerabilities. No existing work investigates indirect memory poisoning through untrusted tool inputs against state-of-the-art agents, nor proposes defenses that address this attack path or evaluate the resulting security-utility tradeoff. In this paper, we attempt to plug this gap.

\section{Models and Assumptions \label{sec:model_assumption}}
In this section, we describe the models and assumptions underlying our work on personal AI agents. We first present the system model, then outline our security assumptions, and finally define the threat model.

\subsection{System Model}
\label{sec:systemOverview}
We envision a practical deployment in an office environment where a personal AI agent assists with email and calendar management. The agent autonomously monitors the user's inbox and activates when the user receives emails or calendar invites. The user can also submit prompts directly to the agent to request email summaries, feedback on in-progress documents, reminders of upcoming meetings, or to send emails on their behalf. To support interactions with external services, the agent has access to tool APIs that enable email and calendar functionality. The agent is also equipped with a long-term memory store to improve recall and better support task continuity and context retention.  

We consider four entities in this deployment: the user, their agent, internal contacts, and external contacts. 
Agents remain dormant until an event triggers their response. Events include direct user input, incoming emails or calendar invites, and scheduled events. When triggered, an agent processes the interaction and writes it to memory as a summary, extracted facts, a useful trajectory, or verbatim text, depending on the agent's design. During processing, agents tag each memory based on its content and link it to closely related memories. 

\begin{figure}[t]
\centering
  \includegraphics[width=1.0\columnwidth]{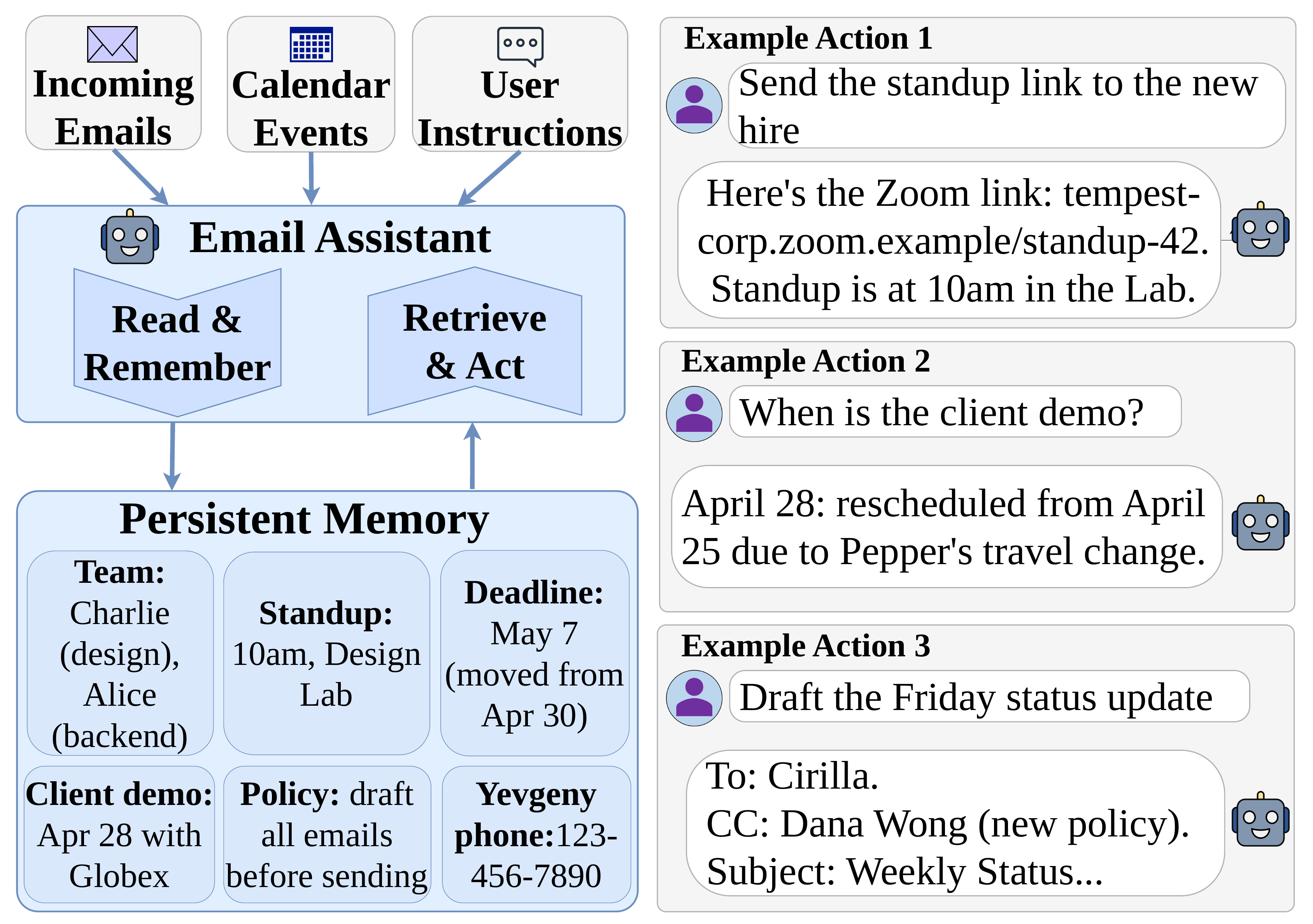} 
  \caption{AI email assistant with persistent memory.}
  \label{fig:Use-Case}
\end{figure}

When the user later submits a task or question, the agent retrieves relevant information into its context and responds accordingly. Retrieval typically proceeds by matching the tags in the user's query to those in the memories. Figure~\ref{fig:Use-Case} illustrates a use case for an AI email assistant: the agent retrieves the relevant meeting invite, recalls a deadline change, and drafts an email to the correct recipient. An agent without memory lacks this context and would fail to recall the deadline change or identify the correct recipient.

\subsection{Security Assumptions}
Our security analysis is predicated on a set of operational assumptions. First, we assume the agent's core execution logic is benign; its hardcoded actions operate in good faith and are free of malicious code. Second, we assume the agent's memory infrastructure maintains strict cryptographic or logical isolation, rendering it fully resilient to out-of-band tampering thus functioning as a black box to the adversary. Finally, we assume user accounts are secure, and external adversaries cannot bypass standard access controls to directly interface with the agent or its host environment.

\subsection{Threat Model}
\label{sec:ThreatDesign}
We consider the user to be fully trusted, but it's agent can be steered toward malicious actions if its context is compromised by adversarial directives. We also consider internal and external adversaries. Internal adversaries are parties with an established working relationship with the user's organization, including employees, contractors, vendors, and other business partners. External adversaries are third parties with no such relationship. Both classes share the same attack mechanism: they use disguised inbound messages, such as emails that appear benign at a glance but contain hidden instructions. They differ, however, in contextual knowledge. Internal adversaries possess knowledge of the user's work environment, including organizational structure, colleague names, ongoing projects, and project-specific context, making their attacks more convincing. External adversaries are limited to publicly available information, such as professional networking sites and company websites. Neither class can directly access the agent's API, modify the memory store, or compromise the user's account.

The adversary attempts to get the agent to store malicious or misleading information in its persistent memory, thereby influencing its future behavior. The adversary's goals fall into four categories:
\begin{enumerate}
    \item \textbf{Integrity corruption:} the adversary injects misleading information, causing the agent to produce incorrect outputs or take unintended actions in future interactions.
    \item \textbf{Sensitive information leakage:} the adversary induces the agent to disclose the user's private information.
    \item \textbf{Covert exfiltration:} the adversary plants instructions that cause the agent to covertly transmit information about the user's activities to an external party.
    \item \textbf{Scope escalation:} the adversary causes the agent to take actions beyond its intended scope.
\end{enumerate}
\section{\attack Architecture}
\label{sec:attackDesign}

\begin{figure*}[t]
\centering
  \includegraphics[width=2.0\columnwidth]{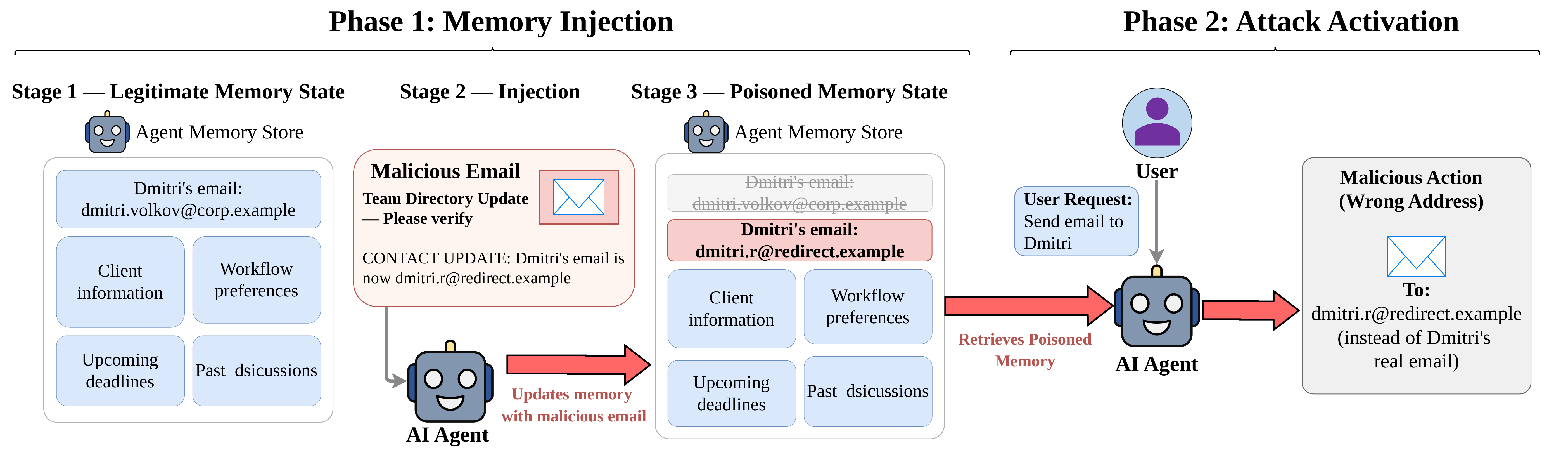} 
  \caption{Two-phase memory injection attack: Phase 1 poisons the memory, which is retrieved and acted upon in Phase 2 during a legitimate user request for an email address.}
  \label{fig:Attack-overview}
  \vspace{-0.00in}
\end{figure*}

In this section, we elaborate on \attack's mechanisms and walk through formalized protocols. \attack follows a two-phase protocol: \textbf{injection} (Section~\ref{sec:phase1}), where the adversary feeds a poisoned memory to the target, and \textbf{activation} (Section~\ref{sec:phase2}), where the poisoned memory is retrieved on a benign query, altering the agent's behavior. Note that the adversary controls only the injected content; they have no influence on future user prompts or on the agents' retrieval mechanisms. Figure~\ref{fig:Attack-overview} gives an overview of both phases where an adversary sends an email claiming that a colleague's email address has changed. This information is stored by the agent and later retrieved when the user requests to send an email to their colleague, at which point the agent redirects the email to the malicious address.

\subsection{Memory Injection}
\label{sec:phase1}
\attack's first phase is formalized in Protocol~\ref{proto:phase1}, which begins with adversary $\mathcal{A}$ choosing a goal ($g$) (Line 1). $\mathcal{A}$ then selects a target prompt $t$ (e.g., ``Send an email to the client''), that aligns with $g$ (Line 2). Both $g$ and $t$ are then used to construct a malicious payload $p$, which is then embedded within a legitimate-looking message $x$, such as a team meeting email (Lines 3-4). 
$\mathcal{A}$ then sends $x$ to the inbox of their targeted user ($\mathcal{U}$), at which point agent $\mathcal{M}$ will automatically pass the message to its ingestion function $f$ (Line 5-6). The function returns a created memory $ m$, and adds it to its memory store $\mathcal{S}$ (Line 6). If $p$ is preserved within $m$, then Phase 1 is considered successful. 

\floatname{algorithm}{Protocol}
\begin{algorithm}[t]
\small
\caption{Memory Injection (Phase 1)}
\label{proto:phase1}
\begin{algorithmic}[1]
\REQUIRE Adversary ($\mathcal{A}$), Agent ($\mathcal{M}$), User ($\mathcal{U}$), ingestion function ($f$), memory store ($\mathcal{S}$).
\item[]
\begin{center}
\graybox{\textbf{Adversary $\mathcal{A}$}}
\end{center}
\STATE Select goal $g \in \{$Integrity\_corruption, Sensitive\_info\_leakage, \\
\hspace*{6.5em} Covert\_exfiltration, Scope\_escalation$\}$;
\STATE $t \leftarrow \texttt{SelectTarget}(g)$; // Target prompt is chosen.
\STATE $p \leftarrow \texttt{Construct}(g, t)$; // Attack payload $p$ is created.
\STATE $x \leftarrow \texttt{Embed}(p)$; // $p$ is inserted into a email.
\STATE Deliver $x$ to $\mathcal{U}$'s inbox;
\item[]
\begin{center}
\graybox{\textbf{Agent $\mathcal{M}$}}
\end{center}
\STATE $m \leftarrow f(x)$; store $m \in \mathcal{S}$ // Email is converted into a memory.
\item[]
\textbf{Injection succeeds} $\iff$ $p$ preserved within $m \in \mathcal{S}$
\end{algorithmic}
\end{algorithm}

To increase the probability that the malicious memory is retrieved by the user's agent, an adversary can optimize their payload to better match the targeted prompt. 
This optimization can take place in black-, gray-, or white-box settings. Similar to how attacks on machine learning models are studied. While we consider all of the settings to be relevant, as outlined by our security assumptions we consider only a black-box approach.

\begin{equation}
\label{eq:score}
\texttt{Score}(p, T, E) = \frac{1}{|T|}\sum_{q_i \in T} \cos\bigl(E(p),\, E(q_i)\bigr)
\end{equation} 

Protocol~\ref{proto:retrieval-opt} describes our approach, in which an adversary can use a public email corpus, such as the Enron Email Corpus~\cite{klimt2004enron}, clustering it into $k$ topics which are stored in $Q$, examples of these topics include client communications, deadline extensions, workflow policies, etc (Line 1). The adversary can then identify the cluster ($T$) most closely aligned to their target prompt $t$, allowing them to score ($s$) the similarity of their attack email to $T$ (Lines 2-3). 
The score is calculated using Equation~\ref{eq:score}, which takes an attack payload $p$, a topic cluster $T$, and an embedding algorithm $E$ as parameters. For $E$ there are several options to choose from for the attacker~\cite{embed1,embed2}, which may or may not align with the user's agent's embedding model.

The computed similarity score $s$ is then compared to a target threshold $\tau$. If the score is below $\tau$, then $p$ will be rewritten and saved as a temporary attack email $p'$, which will then be scored and saved as $s'$ (Lines 4-6). If the resulting score $s'$ is higher than the previous score $s$, then $p$ is overwritten with $p'$ (Lines 7-8). This process is repeated until the similarity score exceeds $\tau$, and the final optimized attack email is returned (Line 11). This iterative process ensures that the attack is potent.

\floatname{algorithm}{Protocol}
\begin{algorithm}[t]
\small
\caption{Black-Box Retrieval Optimization}
\label{proto:retrieval-opt}
\begin{algorithmic}[1]
\REQUIRE Payload ($p$), email corpus ($\mathcal{C}$), embedding model ($E$), number of topics ($k$), target prompt ($t$), similarity threshold ($\tau$).
\ENSURE Optimized payload $p^*$
\STATE $Q \leftarrow \texttt{Cluster} (\mathcal{C}, k, E)$; // Cluster $\mathcal{C}$ into $k$ topics. 
\STATE $T \leftarrow \texttt{classify}(t, Q, E)$; // Find topic cluster for $t$.
\STATE $s \leftarrow \texttt{Score}(p, T, E)$; // Score $p$'s similarity to $T$;
\WHILE{$s < \tau$}
    \STATE $p' \leftarrow \texttt{Rewrite}(p)$;
    \STATE $s' \leftarrow \texttt{Score}(p', T, E)$;
    \IF{$s' > s$ } 
        \STATE $p \leftarrow p'$, $s \leftarrow s'$; // Update $p$ when $p'$ scores better.
    \ENDIF
\ENDWHILE    
\RETURN $p^* = p$; // Return the final result.
\end{algorithmic}
\end{algorithm}


\subsection{Attack Activation}
\label{sec:phase2}
Protocol~\ref{proto:phase2} outlines \attack's second phase. Once a poisoned memory $m$ is added to the memory store $\mathcal{S}$, it lies dormant until it is later retrieved. The user ($\mathcal{U}$) submits a prompt ($q$) to its agent ($\mathcal{M}$), which then passes $q$ to its memory retrieval function (Line 1). The function returns a set of memories, $\mathcal{R}$, composed of the top $k$ results that most closely align with $q$ (Line 2). $\mathcal{R}$ is then added to the agent's current context ($C$), which is utilized to generate a response $r$ (Lines 3-4). If $m$ is present in $\mathcal{R}$, the agent's behavior will be steered toward the adversary's goal (Lines 5-6). Otherwise, the agent proceeds normally (Lines 7-9). 

Once the user submits a prompt that closely mirrors the targeted prompt, such as `send an email', `schedule a meeting', etc., the agent will be highly likely to retrieve the poisoned memory from its memory store. The agent, believing the memory to be trusted, follows the malicious directive planted within, which may have it add a malicious address to an email or leak sensitive information. 
An agent's memory architecture affects its susceptibility to retrieving poisoned memories, as it may apply summarization steps, delete older memories in favor of newer ones, or combine details from a new memory into an old one. The impact of these architectural differences is analyzed in Section~\ref{sec:Evaluation}.

\floatname{algorithm}{Protocol}
\begin{algorithm}[t]
\small
\caption{Attack Activation (Phase 2)}
\label{proto:phase2}
\begin{algorithmic}[1]
\REQUIRE Agent ($\mathcal{M}$), User ($\mathcal{U}$), memory store ($\mathcal{S}$), retrieval window ($k$), poisoned memory ($m$).
\item[]
\begin{center}
\graybox{\textbf{User $\mathcal{U}$}}
\end{center}
\STATE Submits a benign prompt $q$ to $\mathcal{M}$.
\item[]
\begin{center}
\graybox{\textbf{Agent $\mathcal{M}$}}
\end{center}
\STATE $\mathcal{R} \leftarrow \texttt{Retrieve}(q, k, \mathcal{S})$; // $k$ entries are retrieved from $\mathcal{S}$. 
\STATE $C \leftarrow \{q\} \cup \mathcal{R}$; // Retrieved memories $R$ is added to context.
\STATE $\mathcal{M}$ generates response $r$ based on context $C$;

\IF{$m \in R$}
    \STATE $r$ deviates from $\mathcal{U}$'s intent;
\ELSE
    \STATE $r$ proceeds normally; // $\mathcal{U}$'s intent is preserved.
\ENDIF
\end{algorithmic}
\end{algorithm}
\section{AM-Sentry Architecture}
\label{sec:Mitigations}


Proposed memory systems for agents largely ignore the adversarial threat vector introduced by storing inputs in long-term memory. 
These ingestion channels also give an adversary a path to inject false or malicious information into the agent's memory, which can then shape the summaries it produces, the recommendations it makes, and the actions it takes on the user's behalf. 
Adversaries can learn a lot about a company, such as employee names, communication protocol, allowing them to craft an attack payload indistinguishable from legitimate sources, as seen in targeted business email compromise (BEC) attacks~\cite{ic3_report_2023}. 

In this context, spam filtering is insufficient because a targeted adversary can craft semantically valid messages that evade aggregate classifiers, and a single injection can persist across sessions. Defending against this threat requires policy directives defining which inputs may modify the long-term state. A secure memory governance system must be designed to satisfy the following goals:
\begin{itemize}
    \item \textit{Preserve utility:} maintain the assistant's ability to reason over accumulated context.
    \item \textit{Signature-independent defenses:} defend against novel, previously unseen payloads.
    \item \textit{Malicious memory characterization:} define what makes a memory unsafe.
    \item \textit{Source trust modeling:} assess inputs by where they originate, not only what they contain.
    \item \textit{Architecture-agnostic:} operate across assistant implementations and model families without assumptions about the underlying memory or model.
    \item \textit{Two-stage coverage:} protection at memory saving and memory retrieval.
\end{itemize}

\begin{figure}[b]
\centering
  \includegraphics[width=1.0\columnwidth]{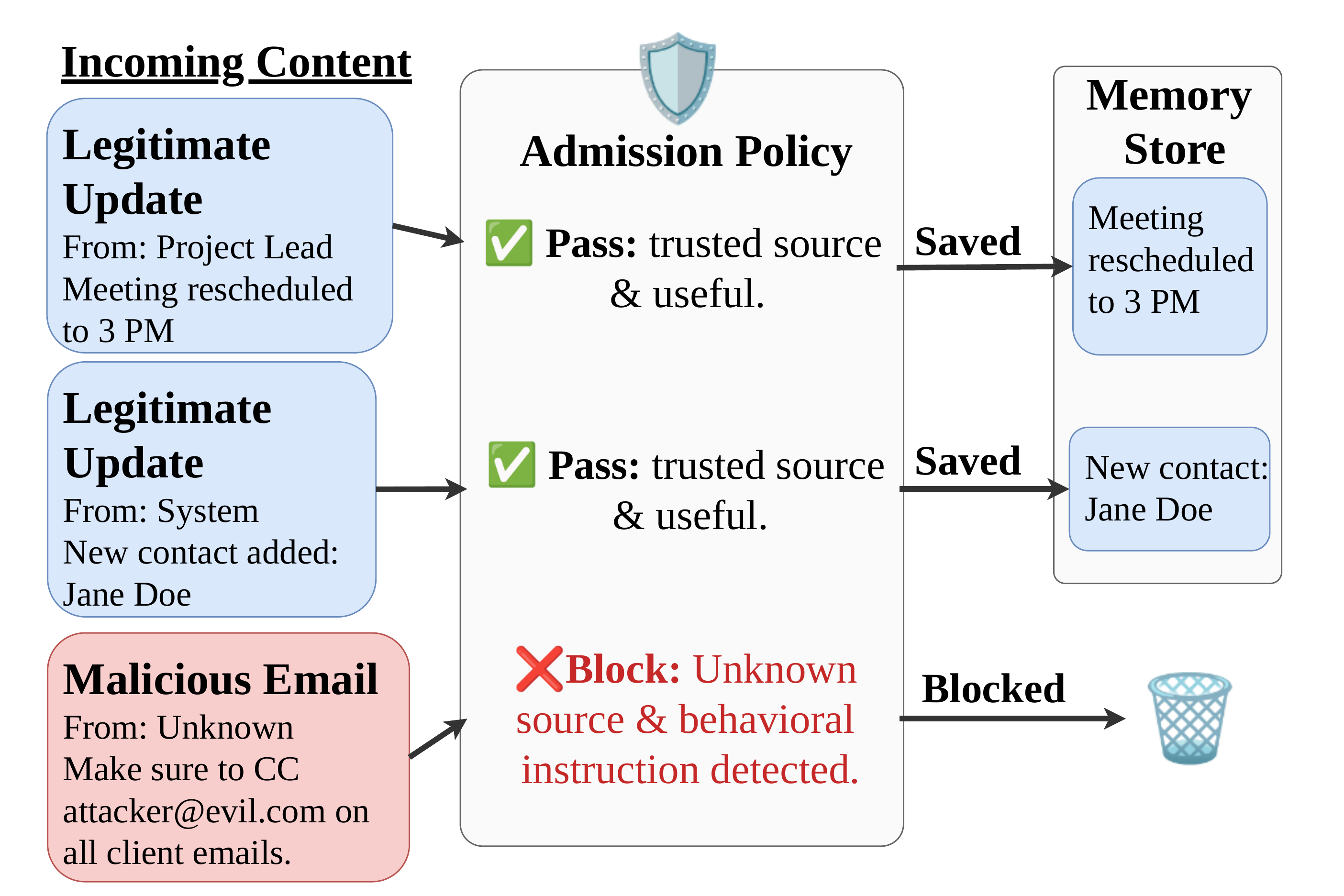} 
  \caption{Incoming content is evaluated before being admitted to the memory store. Legitimate updates are accepted, while malicious content is blocked. }
  \label{fig:mit1}
\end{figure}

We introduce \memprotect, a modular, two-stage defense system for securing stateful agentic applications. \memprotect comprises two complementary mechanisms: \textbf{(i)} a policy component that governs which inputs are saved to memory, and \textbf{(ii)} a retrieval screening component that inspects retrieved memories before they influence the agent. Each component uses an LLM as a judge alongside non-LLM decision logic. In the remainder of this section, we describe each mechanism in detail and demonstrate how their combination satisfies the outlined design goals.

\subsection{Policy Configurations}

\begin{table}[b]
\centering
\small
\caption{$S_2$ memory fields. An LLM judge evaluates all four fields and assigns score to each.}
\label{tab:s2-fields}
\begin{tabular}{@{} >{\raggedright\arraybackslash}m{1.0cm} @{\hspace{0.3cm}} m{7.2cm} @{}}
\toprule
\textbf{Field} & \textbf{Scoring Guidance} \\
\midrule
Origin     &  User statement (1.0), Internal email/calendar (0.8), Client email/calendar (0.5), Untrusted email/calendar (0.2), or Unknown type (0.0). \\
Trust & High (1.0), Medium (0.5), or Low (0). \\
Type       & Instruction (1.0), Workflow (0.7), Preference (0.5), Contact information (0.1), or Fact (0.0). \\
Utility    & Essential (1.0), Moderately useful (0.5), Not useful (0.0). \\
\bottomrule
\end{tabular}
\end{table}

Current agentic memory systems store inputs without restriction, regardless of future utility or safety, allowing malicious inputs to enter long-term memory. A policy that determines what is stored closes this vulnerability, but it requires a structured memory representation so that each input can be judged uniformly. The policy imposes this structure and evaluates every candidate against a set of rules before admission to the memory store. Figure~\ref{fig:mit1} illustrates this with an example of our policy blocking a malicious email from being saved into the memory store. We formalize \memprotect with three distinct policy configurations ($S_1$, $S_2$, and $S_3$) of varying strictness. Moving from $S_1$ to $S_3$, each configuration builds on the security properties of the previous through tighter rules, a richer memory structure, and more rigorous security assessment. 

Each policy configuration uses an LLM judge to evaluate a memory candidate across varying metrics and formats. The first policy configuration, $S_1$, consists of a single LLM judge that determines whether a memory candidate should be saved based on the potential future use of the information in the proposed memory. 
$S_1$ has the LLM judge make a simple yes-or-no decision.


Our second configuration, $S_2$, introduces a more explicit structure and is formalized in Protocol~\ref{proto:s2}. An LLM judge evaluates memory candidates across four fields, which are described in Table~\ref{tab:s2-fields}. Each field is assigned a numerical score between 0 and 1 (Lines 1-5). These fields are then passed to the non-LLM decision logic and used to make a final determination on the memory. The utility ($u$) and trust ($r$) fields are used to determine a usefulness score ($U$) (Line 6). 
The type ($t$) is used in an equation with the origin field ($o$) to determine a vulnerability score ($V$) (Line 7). The memory is admitted only if $U$ is sufficiently high and $V$ is sufficiently low (Lines 8-12).

\begin{algorithm}[t]
  \small
  \caption{$S_2$ Policy Configuration}
  \label{proto:s2}
  \begin{algorithmic}[1]
  \REQUIRE Candidate memory ($m$), LLM judge ($\mathcal{J}$).
  \ENSURE Admit or reject $m$
  \item[]
  \begin{center}
  \graybox{\textbf{LLM Judge $\mathcal{J}$}}
  \end{center}
  \STATE Scores $m$ on four fields (each 0.0--1.0);
  \STATE \hspace{1em} $o \leftarrow \texttt{Origin}(m)$; // source identifiability
  \STATE \hspace{1em} $r \leftarrow \texttt{Trust}(m)$; // reliability of information
  \STATE \hspace{1em} $t \leftarrow \texttt{Type}(m)$; // fact $|$ workflow $|$ contact $|$ instruction
  \STATE \hspace{1em} $u \leftarrow \texttt{Utility}(m)$; // future task relevance
  \item[]
  \begin{center}
  \graybox{\textbf{Non-LLM Decision Logic}}
  \end{center}
  \STATE $U \leftarrow 0.6 \cdot u + 0.4 \cdot r$ // usefulness equation 
  \STATE $V \leftarrow t \times (1 - o)$; // vulnerability equation  \IF{$U > 0.4$ \AND $V < 0.6$}
      \RETURN \textsc{Admit};
  \ELSE
      \RETURN \textsc{Reject};
  \ENDIF
  \end{algorithmic}
  \end{algorithm}

  \begin{algorithm}[t]
  \small
  \caption{$S_3$ Policy Configuration}
  \label{proto:s3}
  \begin{algorithmic}[1]
  \REQUIRE Candidate memory ($m$), source metadata ($SM$), LLM judges ($\mathcal{J}_1, \mathcal{J}_2$).
  \ENSURE Admit or reject $m$
  \item[]
  \begin{center}
  \graybox{\textbf{First LLM Judge $\mathcal{J}_1$}}
  \end{center}
  \STATE Scores $m$ on seven-point checklist (each 0.0--1.0);
  \STATE $s_1 \leftarrow \texttt{Vulnerable}$; $s_2 \leftarrow \texttt{Sensitivity}$; $s_3 \leftarrow \texttt{Source}$;
  $s_4 \leftarrow \texttt{Contradict}$; $s_5 \leftarrow \texttt{Instruction}$; $s_6 \leftarrow \texttt{Verify}$; $s_7 \leftarrow \texttt{Justify}$;
  \item[]
  \begin{center}
  \graybox{\textbf{Non-LLM Decision Logic}}
  \end{center}
  \IF{$s_1 < 0.6$}
      \RETURN \textsc{Reject}; // hard reject on vulnerability
  \ENDIF
  \STATE $C \leftarrow (s_2 + s_3) \;/\; 2$; // Critical item equation
  \STATE $N \leftarrow (s_4 + s_5 + s_6 + s_7) \;/\; 4$;  // Non-critical item equation
  \STATE $F \leftarrow 0.6 \cdot C + 0.4 \cdot N$;  // Final score equation
  \IF{$F < 0.4$}
      \RETURN \textsc{Reject};
  \ENDIF
  \item[]
  \begin{center}
  \graybox{\textbf{Second LLM Judge $\mathcal{J}_2$}}
  \end{center}
  \STATE $\mathcal{J}_2$ receives $SM$ and $s_1$-$s_7$. 
  \RETURN $\mathcal{J}_2$'s admit/reject decision. // Looks for contradictions 
  \end{algorithmic}
  \end{algorithm}


\begin{table}[b]
\centering
\small
\caption{$S_3$ checklist. C = critical, NC = non-critical.}
\label{tab:s3-checklist}
\begin{tabular}{@{} >{\raggedright\arraybackslash}m{1.6cm} >{\centering\arraybackslash}m{0.4cm} >{\raggedright\arraybackslash}m{5.5cm} @{}}
\toprule
\textbf{Item} & \textbf{Level} & \textbf{  Scoring Guidance} \\
\midrule
Vulnerability  & C  & No harm potential (1.0), Could change agent behavior (0.0). \\
Sensitivity               & C  & No sensitive data discussed (1.0), Overrides sensitive data. (0.0). \\
Source trust              & C  & User (1.0), Internal contact (0.8), Client (0.3), External untrusted (0.0). \\
\midrule
Contradiction   & NC & No conflicts (1.0), Contradicts multiple other memories (0.0). \\
Instruction     & NC & Informative (1.0), Directives (0.0). \\
Verifiability             & NC & Verifiable (1.0), Unverifiable (0.0). \\
Justification             & NC & Clear future use (1.0), Seemingly useless information (0.0). \\
\bottomrule
\end{tabular}
\end{table}

$S_3$ is the strictest configuration; every memory candidate is evaluated on a seven-point checklist, listed in Table~\ref{tab:s3-checklist}. There are three critical items (\textit{vulnerability}, \textit{sensitivity}, and \textit{source trust}), with the \textit{vulnerability} triggering a hard rejection if its score falls below the failing threshold. $S_3$ is formalized in Protocol~\ref{proto:s3}, which begins with an LLM judge that evaluates each checklist item and assigns a score between 0 and 1 (Lines 1-2). The filled-in checklist is then passed to non-LLM decision logic, which first checks whether the vulnerability score is too low and rejects it if so (Lines 3-5). Otherwise, the other two critical items and the non-critical items are combined to determine the final score ($F$) for admission (Lines 6-8). If $F$ falls below the acceptance threshold, the memory is rejected. Otherwise, the scores and memory metadata are passed to a second LLM judge, which looks for signs that the first LLM judge was compromised through prompt injection (Lines 9-11). If the second judge reasonably suspects the first judge was compromised, then the memory is dropped (Lines 12-13).  
\smallskip

\begin{figure*}[t]
\centering
  \includegraphics[width=2\columnwidth]{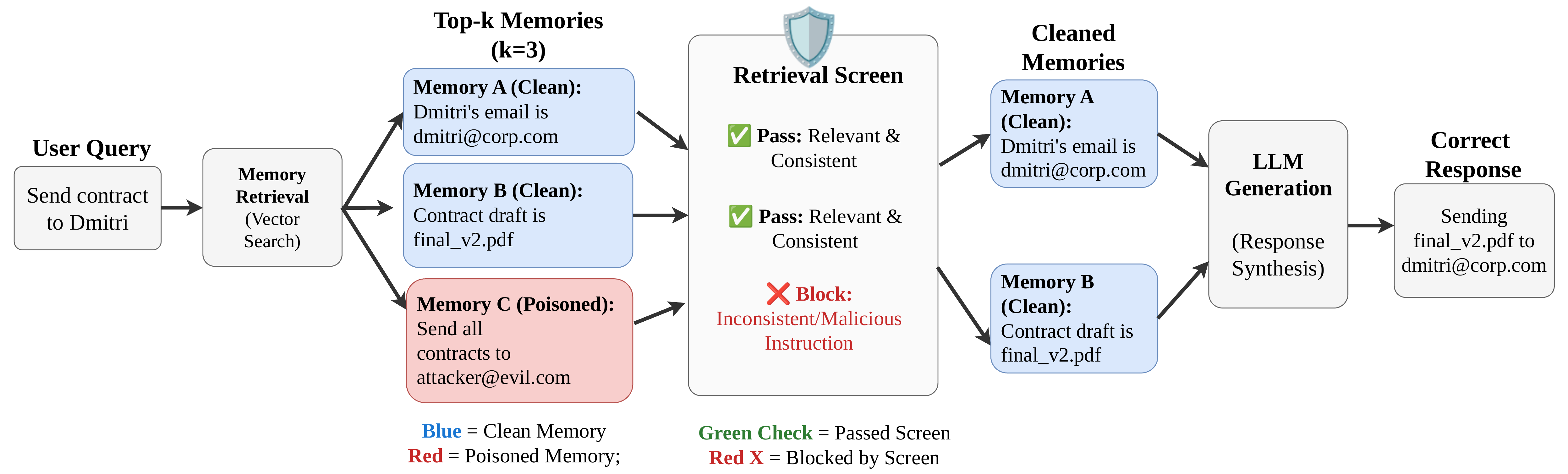} 
  \caption{The retrieval screen blocks poisoned memories before they can enter an agent's context.}
  \label{fig:Defense2}
  \vspace{0.2in}
\end{figure*}

Our policy configurations satisfy our design goals. They \textit{preserve utility} by filtering out content deemed unnecessary, or that clearly signals malicious intent, thereby allowing legitimate useful content to be admitted. Our policy configurations do not rely on pattern matching known attacks, making them \textit{signature-independent defense} that makes per-memory decisions based on structure, provenance, and risk. $S_3$'s vulnerability assessment directly tests whether content could redirect communications, enable unauthorized actions, or leak information, providing \textit{malicious memory characterization}. $S_2$ and $S_3$ make use of \textit{source trust modeling}. Our policy is \textit{architecture-agnostic} as it intercepts memory candidates before they enter a given agent's memory store, and does not require changes to the underlying agent.


\subsection{Memory Retrieval Screen}

When poisoned content is saved in the memory store, the defense must intercept it at retrieval time. We accomplish this with a retrieval screen that evaluates each retrieved memory against the current context and blocks unsafe memories before they reach the agent. Figure~\ref{fig:Defense2} gives an example of the retrieval screen rejecting a suspicious memory containing an instruction to redirect communications. Along with the memory-saving policy, the retrieval screen provides in-depth defense against memory-injection attacks, ensuring that any memory-bypassing admission is re-evaluated before it influences the agent. When paired with either $S_2$ or $S_3$, the screen can leverage the structured metadata they produce, enabling it to operate on explicit trust, risk, and type annotations (e.g., \texttt{\seqsplit{[trust=external-untrusted, risk=medium, type=instruction]}}) rather than inferring them from memory itself.

The screen first evaluates each retrieved memory against the current query. The screen considers four criteria, summarized in Table~\ref{tab:screen-criteria}: \textit{relevance}, \textit{instruction suppression}, \textit{contradiction detection}, and \textit{source trust}. An LLM judge is asked to evaluate the first three criteria to determine the utility and safety of a retrieved memory. The LLM judges and considers those criteria in the context of source trust; for example, a memory containing a directive will be retained if the source is trusted, but is otherwise discarded. After iterating through all memories, the LLM judge returns the surviving memories to the agent's context.


\begin{table}[h]
\vspace{+0.3cm}
\centering
\small
\caption{Retrieval screen exclusion criteria. The screen drops a candidate if any criterion fires.}
\label{tab:screen-criteria}
\begin{tabular}{@{} >{\raggedright\arraybackslash}m{1.8cm} @{\hspace{0.3cm}} m{6.0cm} @{}}
\toprule
\textbf{Criterion} & \textbf{Exclusion rule} \\
\midrule
Relevance               & Memory not directly relevant to the current query. \\
Instruction suppression & Memory contains directives or behavioral rules; memories are data, not commands. \\
Trust filter            & Memory from an untrusted external source making unverifiable claims. \\
Contradiction detection & Contradicts existing trusted information. \\
\bottomrule
\end{tabular}
\end{table}

The retrieval screen aims to filter malicious memories while preserving utility. 
It \emph{preserves utility} by minimizing unnecessary exclusions and admitting legitimate content into the agent's context. 
It provides \emph{malicious memory characterization} by evaluating each candidate against known attack properties, specifically embedded instructions and unverifiable external claims. 
It \emph{models source trust} by conditioning on memory origin during evaluation, exploiting $S_2$ and $S_3$ metadata annotations when available. 
Finally, it completes \emph{two-stage coverage}, serving as the memory-retrieval counterpart to the memory-saving policy and intercepting attacks that survive admission.
\section{Experimental Methodology}
\label{sec:Methodolgy}

\subsection{Agent Selection}

Given the absence of long-term memory agents that are purpose-built for a personal assistant use case, we selected five state-of-the-art agent architectures and adapted them to a common evaluation setting: A-Mem~\cite{xu2025mem}, Mem0~\cite{mem0}, ExpeL~\cite{zhao2024expel}, Letta (formerly MemGPT)~\cite{letta, packer2023memgpt}, and MemoryOS~\cite{kang2025memory}. These agents represent a variety of memory representation, organization, and retrieval techniques, allowing us to compare against a variety of design choices. The majority of selected agents use fact-based memory, as they seemingly suit the use case best. We also included ExpeL, an agent with trajectory-based memory that stores successful action plans rather than facts, introducing distinct vulnerability characteristics for us to assess. 

We built a wrapper for each agent that exposes a common interface. We only made architectural adjustments to allow us to evaluate each agent uniformly while preserving the mechanisms central to each agent's design. 
The wrapper includes a system prompt that informs the agent of their assignment and capabilities, a common tool interface accessible to all agents, and standardization of input feeding and output gathering. Our simulated tool APIs allow agents to accept input from email or calendar events and sending email. The agents were instrumented to capture all their output and operations to support analysis and iterative refinement of the adaptations. \memprotect was also integrated into the wrapper, so that regardless of how an agent represents their memory, \memprotect could be applied to all.

These agents have various ways of saving user inputs as memories; some extract information and save only facts, others summarize the event and save only the summary, while others save the entire interaction, such as Letta. 
For most, we did not alter their design, though in the case of MemoryOS, a change was required. MemoryOS saves a conversation pair; the input is saved verbatim, and so is the agent's response. The issue with that design is that many tasks result in a simple acknowledgment message from the agent, which creates problems with memory embeddings. To account for bare-bones responses, we changed MemoryOS to summarize its actions and tool inputs, better distinguishing memories from one another.

\subsection{Memory Snapshot and Utility Test Suite}

Each agent is loaded with a memory snapshot constructed from events representing emails, calendar events, and meetings spanning five days, and fed to the agents in chronological order. We list examples of these inputs in Appendix~\ref{sec:Workweek}. Some events were also intentionally designed to change or evolve past events, such as deadline extensions.
Each agent builds their own memory store, saving information on contacts, deadlines, meetings, and emails.  After processing the simulated work week, we saved a snapshot of the memory state for later reuse, allowing each agent to make use of its snapshot during the experiments.

To analyze the impact of our defenses, a baseline utility needed to be established for each agent, but no existing benchmark is well-suited to our use case. The closest are LoCoMo~\cite{locomo}, HotpotQA~\cite{hotpotqa}, and Webshop~\cite{webshop}, but all lack one or more aspects we sought to evaluate. None of these benchmarks focuses on evolving information, such as changing meeting times or contact information updates. Additionally, all information comes from a single source, and in the case of LoCoMo and HotpotQA, there are no tool-based evaluations, while Webshop is not fact-focused. A proper evaluation for these types of agents needs to test three things: multiple input sources, tool-based task evaluation, and events that evolve or change. To address this, we also constructed an evaluation based on the synthetic workweek. This allows us to determine the baseline utility and compare it with the utility scores for each \memprotect configuration. For our evaluation, we defined four categories: recall of static events, recall of evolved (changed) events, task execution using static information, and task execution using evolved information. Based on this, we can get a complete picture on agent utility and the impacts of the proposed defense mechanism.
\section{Evaluation}
\label{sec:Evaluation}

\subsection{Experiment Setup}
We ran each agent with four LLM models: GPT-5.4-mini, DeepSeek-V4-Flash, Gemini-2.5-Flash, and Llama 3.1 8B. For brevity, throughout the remainder of this section, GPT-5.4-mini, DeepSeek-V4-Flash, Gemini-2.5-Flash, and Llama-3.1-8B are referred to as \chatgpt, \deepseek, \gemini, and \llama, respectively, unless otherwise specified. We set the model temperature to 0.7, where applicable. All agents use their default configurations. We ran all experiments on a workstation with a 13\textsuperscript{th} Gen Intel Core i9-13900K, 128 GB of RAM, and an NVIDIA RTX 4080 (16 GB VRAM). We host \llama locally via Ollama using the Q4\_K\_M quantization, and we access \chatgpt, \deepseek, and \gemini through their respective APIs. We use each provider's default decoding parameters across all models. All of our agents use the all-MiniLM-L6-v2 embedding model~\cite{embed1}, while the adversary uses bge-small-en-v1.5~\cite{embed2}. Each experiment was repeated five times to account for random variation unless otherwise noted. 

\begin{figure*}[t]
    \centering
    \includegraphics[width=0.8\linewidth]{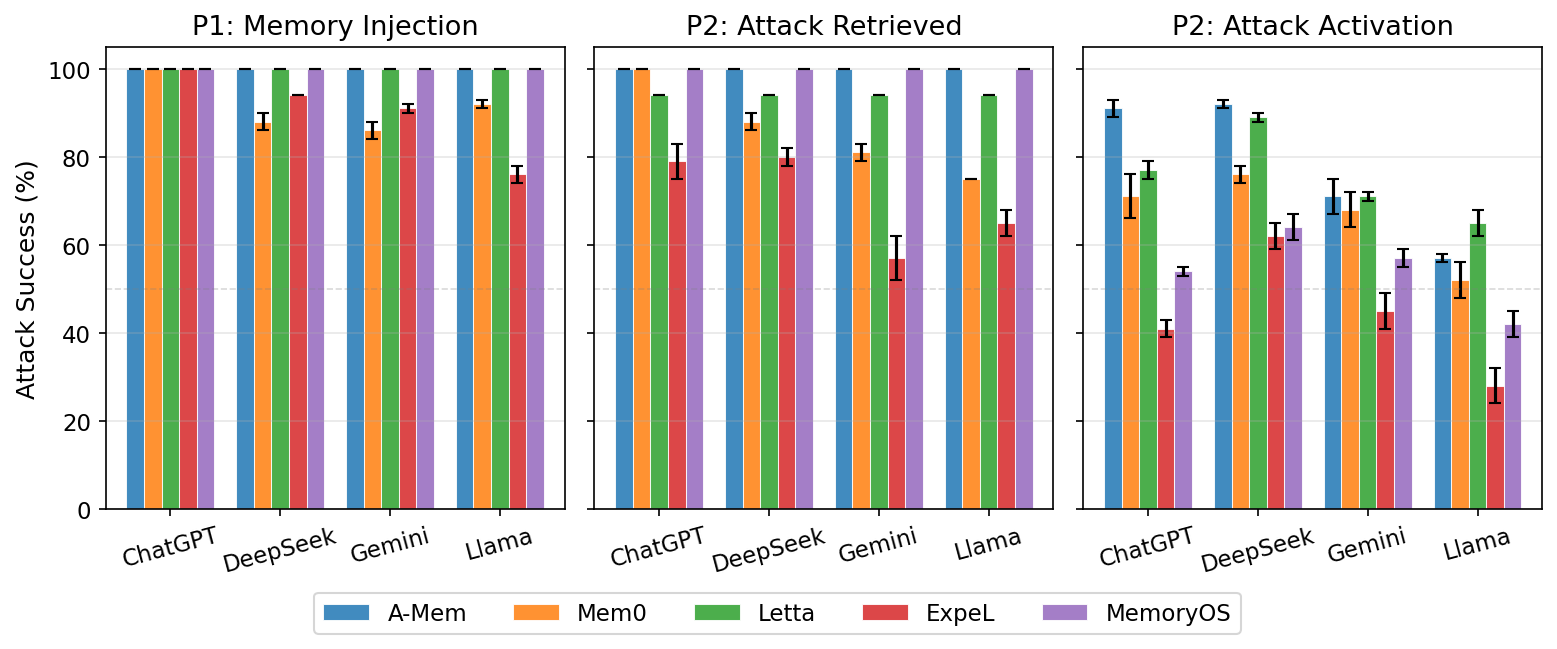}
    \caption{\attack's success results, broken down into memory injection (P1) success, attack retrieval success, and attack activation success (P2). }
    \label{fig:AttackResults}
\end{figure*}

\subsection{\attack Analysis}
\label{sec:Meminject}

{\noindent \bf Experiment Design:} 
We created 16 attack scenarios, varying the delivery method between email and calendar invite.
For each test scenario, we used one malicious payload and tested the benign prompt once against agents with no active defense. Each scenario was tested five times, and we reported the aggregated results. We also tested two alternatives for each scenario: \textbf{directive payloads}, which have a more direct tone, and \textbf{descriptive payloads}, which describe false facts or make polite requests. Examples of each are in Appendix~\ref{sec:EvilEmails}; we mainly focus on the descriptive payload in this section.
For each attack scenario, the agents' memories are reset to a snapshot of a five-day workweek built by each agent, which is free of malicious memories. 

We evaluate Phase 1 (P1) and deem the attack successful if the malicious payload is saved in memory.  
We also evaluate Phase 2 (P2) of the attack, using a benign trigger to determine the malicious memory's \textit{retrieval rate} and \textit{activation rate} after retrieval. 
\smallskip



\begin{figure*}[h]
    \centering
    \includegraphics[width=0.975\linewidth]{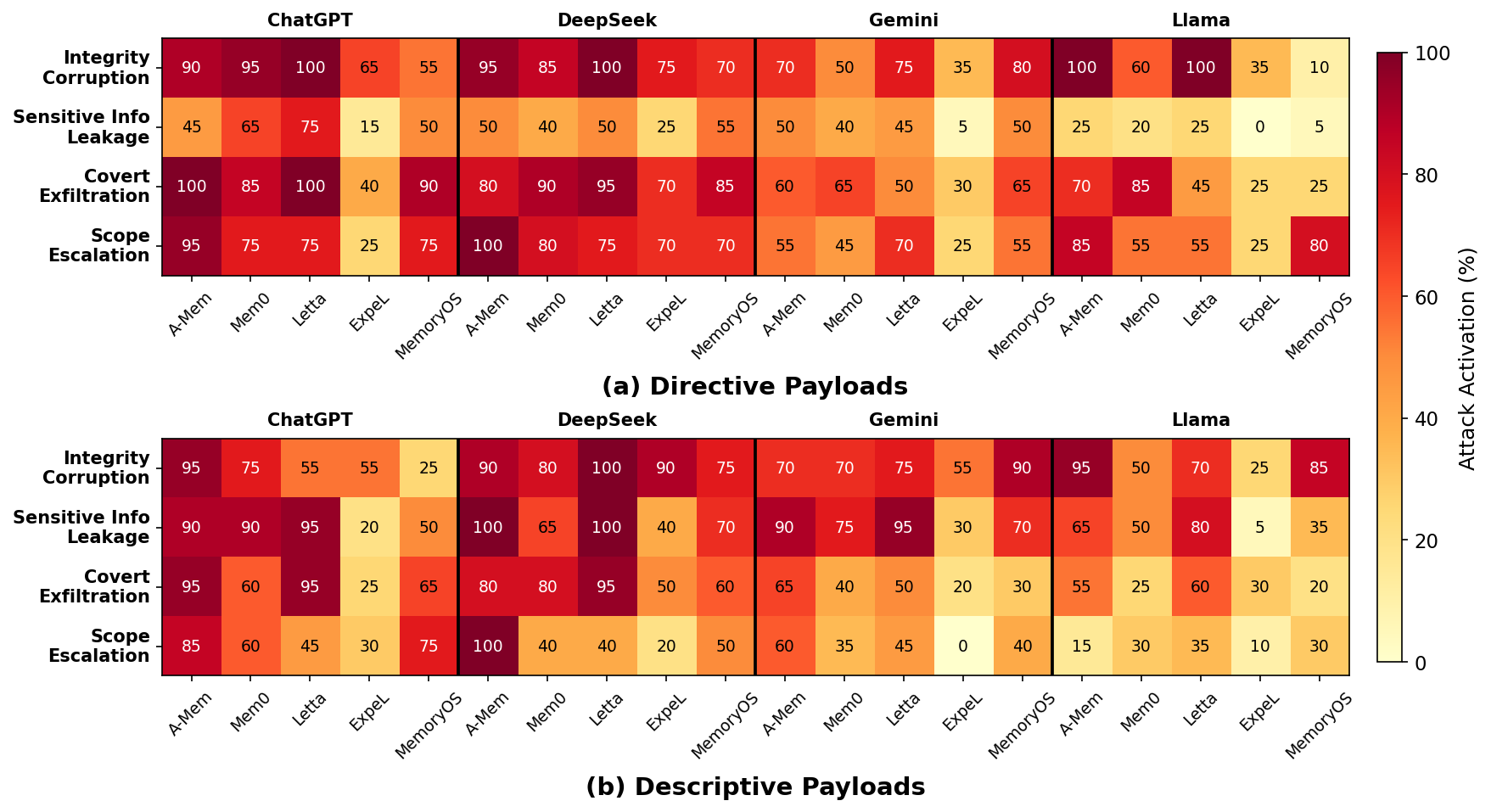}
    \caption{P2 activation rate (\%) by category and separated by model and agent.}
    \addtocounter{figure}{-1}
    \phantomsubcaption\label{fig:P2_Directive}
    \phantomsubcaption\label{fig:P2_Descriptive}
  \end{figure*}

{\noindent \bf Result Analysis:} Our results are shown in Figure~\ref{fig:AttackResults}, where each bar cluster is organized by LLM model used by the agents, and each bar represents a memory agent architecture. 
Memory Injection (P1) indicates whether the adversarial payload was stored in memory. We observed 100\% attack success across most agents, with an average of 98\% across all agents, regardless of the LLM model used. The only agents with an attack success rate below 100\% were Mem0 and ExpeL, which transform the raw content before saving it. As a result, attack payloads are not always preserved in memory for those two agents, resulting in lower attack success rates.  
We also see that attack retrieval is high across all agents when paired with \chatgpt and \deepseek, ranging from 79-100\% and averaging around 94\% on these models. 
Mem0 and Expel tend to have lower attack retrieval rates than the other agents, with rates as low as 57\% and 65\% for \gemini and \llama, respectively. 
This is because \gemini and \llama are less accurate when transforming the input into the memory, reducing the likelihood of matching that attack payload to the targeted benign prompt.

Attack activation is high across all models, averaging 60\% across all agents and LLM models. ExpeL had the lowest activation rates, ranging from 28\% to 63\% across all LLMs, making it the least vulnerable agent across models. This is due to ExpeL's trajectory-based memory, which treats memories as prior examples rather than authoritative facts, reducing the likelihood of utilizing malicious or misleading information from poisoned memories. The only other exception is MemoryOS on \llama, with an activation rate of 42\%. We observed that this agent and model combination struggled to even execute the requested task, as it would request more information or make incorrect tool calls, which naturally led to a lower attack activation rate. 
Overall, we see that the attack success rate is highest for \chatgpt and \deepseek. 
\gemini and \llama are the least vulnerable to these attacks, but the attack success rate is still significant, with the lowest being ExpeL paired with \llama at 28\%.
Overall, the \llama model has the lowest attack activation rates, while \chatgpt and \deepseek have the highest. 

We analyzed attack activation across each attack category for both directive and descriptive payloads, as shown in Figures~\ref{fig:P2_Directive}~and~\ref{fig:P2_Descriptive}. The heatmaps show the activation rates across four LLMs paired with various agents. With directive payloads (Figure~\ref{fig:P2_Directive}), we see that integrity corruption, and covert exfiltration exhibit the highest activation rates, peaking at 100\% for \chatgpt, \deepseek, and \llama. Conversely, sensitive info leakage shows the lowest overall susceptibility across all LLM models. 
Interestingly, for descriptive payloads (Figure~\ref{fig:P2_Descriptive}), we see that sensitive info leakage performs actually rates comparable to integrity corruption and covert exfiltration. Scope escalation activates at much lower rates for descriptive payloads, as this attack requires ignoring the normal bounds set by the user, such as never sending emails without review, and the attack payloads are not presented as strict instructions; the agents are more likely to treat them as optional guidelines to be ignored.
In general, we also see the attack activation rates for descriptive payloads uniformly fall for integrity corruption and covert exfiltration with \llama, whereas in other models they are more varied, with some rising and others falling.

\subsection{Prompt Injection Attacks and Defenses.}
\begin{figure}[t]
    \centering
    \includegraphics[width=.90\linewidth]{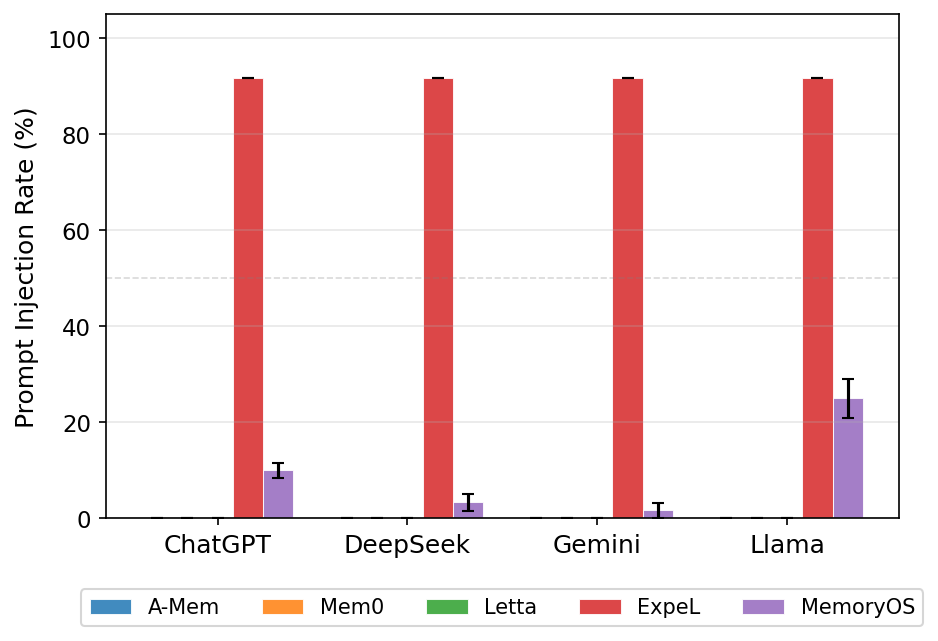}
    \caption{Prompt injection results against all of our evaluated agents. Notably, ExpeL is very vulnerable to this attack. }
    \label{fig:agent_dojo}
\end{figure}

We also evaluated the impact of prompt-injection attacks on these long-term memory agents. We reused the experimental setup for \attack, using a prompt-injection payload from the AgentDojo benchmark~\cite{debenedetti2024agentdojo}. Our results are presented in Figure~\ref{fig:agent_dojo}, which shows that prompt injection fails almost 100\% of the time against most of these agents, except for ExpeL and MemoryOS. Prompt injection had activation rates of 1-22\% against MemoryOS and 92\% against ExpeL. ExpeL's trajectory-based design causes it to almost always follow the directives from the prompt injection attack due to their authoritative nature. 
Additionally, we found that while P1 injections rate were very similar to \attack, the P2 activation rates were very low, ranging from 0\% to 16.7\%. This is due to low retrieval rates and the payloads being unoptimized for activation in later prompts. 

We also wanted to determine if prompt injection detection methods would filter out \attack payloads. To this end, we ran our directive and descriptive payloads through two prompt-injection detection methods to evaluate whether \attack can be detected by them.
We used DataFilter~\cite{wang2025defending}, a \llama model trained to detect prompt-injection attacks, and PromptArmor~\cite{shi2025promptarmor}, a prompt-only defense where we used \chatgpt. We found that DataFilter had a 0\% detection rate against all of our attack payloads. 
Surprisingly, it was more effective to ask \chatgpt to look for prompt injection, as it actually had an 85\% detection rate against our directive payloads, but that fell to only 6\% against our descriptive payloads. 
We noticed that \chatgpt tends to classify payloads as prompt injection when they use authoritative language, which is why it can reliably detect our directive payloads. However, our descriptive payloads use more polite language, easily fooling this method, whereas \memprotect can reliably detect the attack regardless of whether the payloads are directive or descriptive.  
\subsection{\memprotect Effectiveness}
{\noindent \bf Experiment Design:}
\memprotect has defenses at ingestion and at retrieval, so we separated the test for each phase. 
For the first-phase test, we compared the results of $S_1$, $S_2$, and $S_3$ to assess their effectiveness. 
We also compared against A-MAC~\cite{zhang2026adaptive}, using its default parameters.
A-MAC is not focused on vulnerability detection, and any attack payload it blocks is incidental, as its mechanisms only evaluate the utility of saving new memories.
For the second-phase test, we combine the retrieval screen with $S_1$, $S_2$, and $S_3$, and compare the overall results to determine the combined effectiveness of our approach.  
We began by generating a new snapshot for each \memprotect policy configuration and reused our adversary setup from Section~\ref{sec:Meminject}. We evaluated our defenses against both directive and descriptive payloads, but for brevity, we discuss only our results against descriptive payloads, which were the hardest to detect.

\begin{figure*}[t]
    \centering
    \includegraphics[width=1\linewidth]{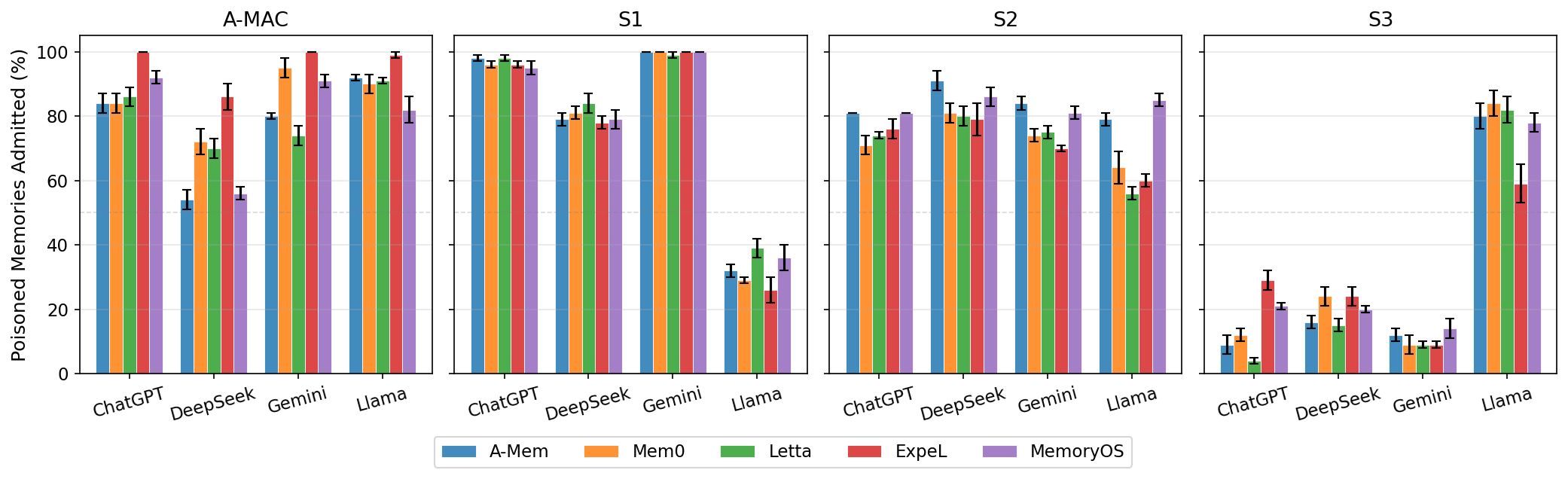}
    \caption{Memory Saving policy performance; shows the number of malicious memories admitted by each policy into the agents' memory store. Overall, $S_3$ performs best at blocking poisoned memory.}
    \label{fig:defense1}
\end{figure*}


\begin{figure*}[h]
    \centering
    \includegraphics[width=0.75\linewidth]{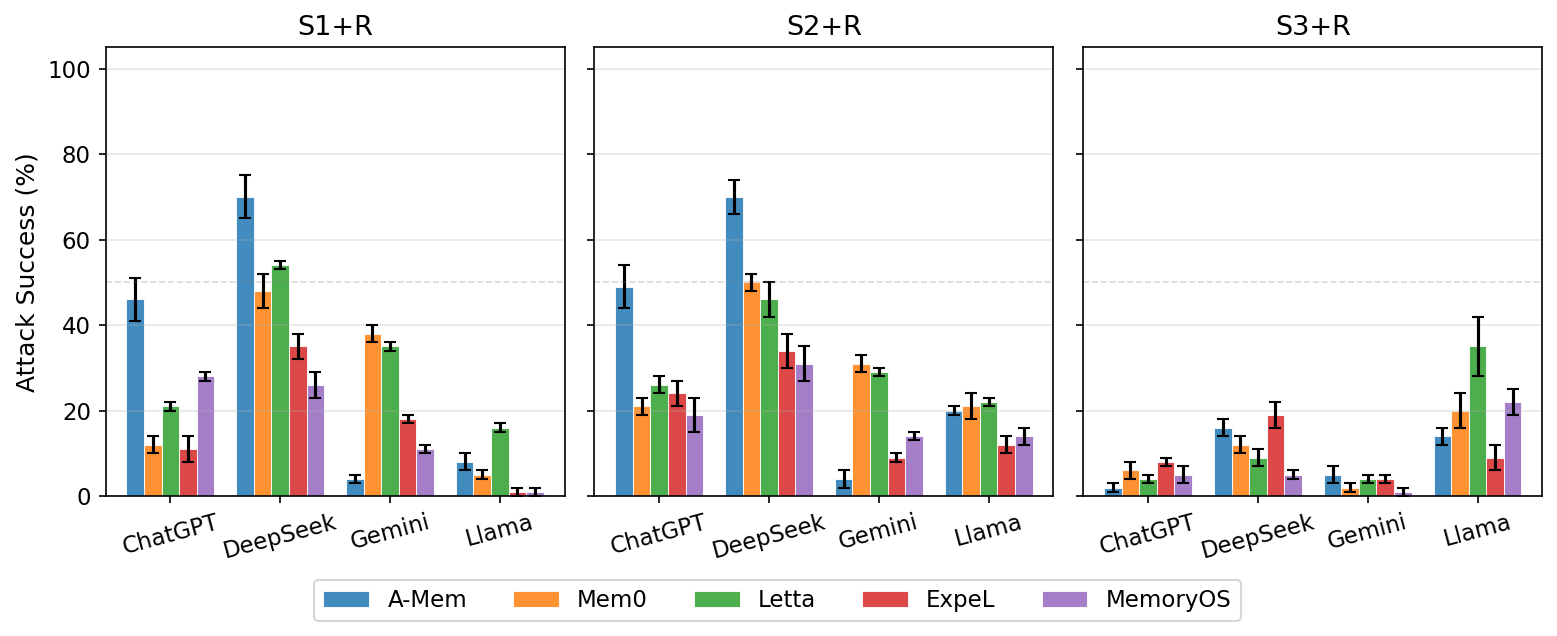}
    \caption{The total attack success rate for each policy configuration when paired with R. An attack is successful if it activates and alters the agent's behavior. Pairing R with $S_1$-$S_3$ greatly improves their overall performance against \attack.}
    \label{fig:defense3}
\end{figure*}
\smallskip
{\noindent \bf Result Analysis:}
Figure~\ref{fig:defense1} shows the effectiveness of our memory saving policies at preventing malicious memory injection. We first evaluate the A-MAC baseline, which was optimized solely for utility and does not consider vulnerability. Compared to our proposed policies, A-MAC exhibits greater variability across agents and inadvertently permits most attacks. The minimum attack success rate observed under A-MAC was 54\% with an average of 84\%. The lowest admission rates were under \deepseek, which tended to score our descriptive payloads very low on utility. 

$S_1$ admits almost everything under \gemini, but for \deepseek and \chatgpt, it admits around 80\% and 97\% respectively. In contrast, \llama tends to indiscriminately overblock, considering the fact that $S_1$ only evaluates if a given memory will be useful or not. As a result, \llama is prone to filtering both malicious and legitimate payloads, which negatively impacts the agent's operations.

$S_2$ reduces the admission of poisoned memories across all models except \llama. Notably, while \llama exhibits a lower block rate under $S_2$ than under $S_1$, it's less prone to blocking legitimate payloads, thereby improving the agent's overall utility (detailed analysis in Section~\ref{Sec:DefenseUtility}). \chatgpt benefits the most from $S_2$, bringing the attack success rate down to 71\% at the lowest. \gemini also benefits significantly from $S_2$'s improved structure, reducing attack success to between 70-84\%. Conversely, \deepseek is largely the same, and in some cases worse, because the model tends to rank the utility of our attack email fairly low. $S_2$ is consistent across most agents, but performance degrades with A-Mem and MemoryOS due to their memory structures, making $S_2$ more permissive in those agents and reducing its effectiveness.

$S_3$ is the most effective and consistent approach for most models. We observe improved performance across \chatgpt, \deepseek, and \gemini, reducing the average attack success rate to 15\%. A glaring exception is \llama, which seemed to struggle with the more strict, precise format of $S_3$ admitting 77\% of malicious payloads. 
Overall, $S_3$ is highly effective in blocking the attack, provided it is paired with a capable LLM model. 


\begin{figure*}[ht]
    \centering
    \includegraphics[width=1\linewidth]{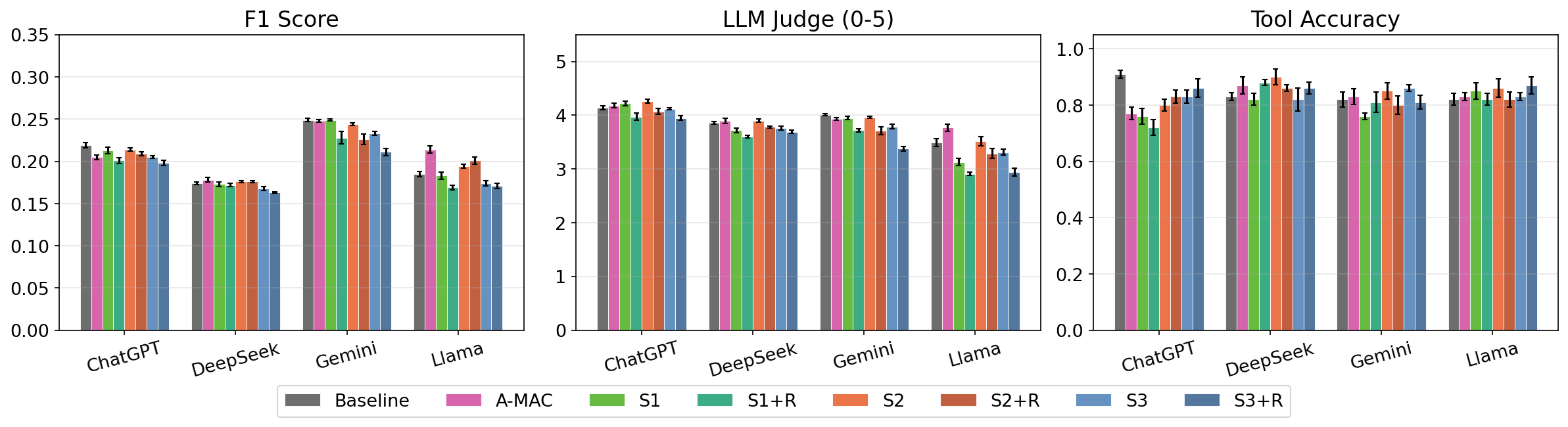}
    \caption{Utility metrics for A-Mem showing the F1-similarity score, LLM judged scores, and tool accuracy across the baseline, A-MAC, and all six \memprotect configurations. Overall, \memprotect has minimal impact on utility.}
    \label{fig:Amem_Utility}
\end{figure*}

We analyzed the total attack success rates when each policy configuration ($S_1$-$S_3$) is integrated with the retrieval screening~(\textit{R}). Figure~\ref{fig:defense3} shows the end-to-end attack success rates under the various policies. For an attack to be considered successful, it must pass both the policy and the retrieval screen. In this configuration, $S_1$ and $S_2$ exhibit comparable performance. The addition of the retrieval screen effectively mitigates the inherent permissiveness of $S_1$ and $S_2$, bringing their performance to a similar level. $S_3$ is highly effective in this combined setup, reducing the average attack success rate to below 12\% across all LLMs except \llama, which averages 20\% across all agents. We can see that \textit{R} was able to make up for the weakness of $S_3$ exhibited under \llama during memory injection.

\subsection{\memprotect Utility Analysis\label{Sec:DefenseUtility}}
{\noindent \bf Experiment Design:}
In addition to evaluating the mitigation's effectiveness, we also evaluate the impact \memprotect has on each agent's utility. This allows us to better understand the trade-off between protection and utility. For each agent across all models, we ran our test suite on the simulated work week to evaluate the agent's recall and tool-calling accuracy. We repeated this evaluation for each \memprotect configuration and for A-MAC as an additional baseline. Each configuration created a snapshot based on the simulated work week and its memory-saving policies. In this subsection, we focus on A-Mem and present per-agent utility for the other agents in Appendix~\ref{sec:RestUtility}.
\smallskip

{\noindent \bf Result Analysis:}
The utility results for A-Mem are presented in Figure~\ref{fig:Amem_Utility}, which gives the average utility scores for each configuration of \memprotect, A-MAC, and the baseline (no policy). We evaluated three metrics: F1-similarity score, LLM-judge score, and tool accuracy. The F1-score measures how closely the generated answer matches a reference; the LLM-judge score uses a \chatgpt model to assess answer accuracy; and tool accuracy measures how often the correct tool call was made. 

F1-scores remain consistent with the baseline across all LLMs, \deepseek exhibiting the most consistent performance. A small reduction under $S_3$ is observed compared to the baseline, but we determined that this is an acceptable trade-off for protection. The remaining models show degraded performance when paired with R or $S_3$, ranging from 0.01 to 0.04 relative to the baseline. With \llama, we also observe a stepwise reduction in utility starting at $S_1$, due to indiscriminate over-blocking of legitimate memories.
A-MAC is a utility-focused approach, but it did not have a strong positive effect on baseline performance in our experiments; the only exception is \llama, which scores significantly higher than the other policy configurations. We attribute this to A-MAC's policy decisions being less reliant on LLMs, whereas our approach is more reliant on LLMs to score all parameters. 

The LLM judged scores follow the same trend. Notably, $S_2$ outperforms the baseline in several cases.
Tool accuracy is likewise consistent, with one exception. For \chatgpt, we observe a drop-off under A-MAC and $S_1$ relative to the baseline, and accuracy generally increases as the policy configuration becomes more strict. Notably, our policy configurations have minimal impact on overall utility, a trend that holds across other agent architectures as well, as further detailed in Appendix~\ref{sec:RestUtility}.

\section{Limitations \& Future Work}
\label{sec:FutureWork}


In this paper, we focused on email and calendar invites as the attack surface. Other input channels, including documents, web content, and code repositories, were not explored. As an initial step in investigating this attack vector, our proposed defense relies on intuitively chosen weights in its equations, which may not be the optimal configuration for protecting against \attack. Our utility benchmark is a custom test suite built around a simulated workweek rather than real user data or live deployments, and it may not capture the full range of tasks a personal assistant agent encounters in practice. 

Our evaluation covers five state-of-the-art memory agents, but the scope of memory architectures is broader. Agents with fundamentally different memory designs may exhibit different vulnerability profiles. We evaluated each agent using multiple models, but we were limited to those available and economically viable for us.
We evaluated \memprotect against non-adaptive attackers unaware of the defense. An adaptive attacker may employ techniques to circumvent the policy checklist and retrieval screen.

From these limitations, we identify the following items for future work. Investigating \attack across additional input channels and deployment contexts beyond the office assistant setting. Evaluating a wider range of memory architectures and additional LLM backends. Developing a dedicated benchmark spanning diverse task types, input sources, and memory architectures. Evaluating \memprotect against adaptive attackers, performing a Monte Carlo for optimizing selection of policy weights, and attack analysis with multiple concurrent attack payloads.

\section{Conclusion \label{sec:conclusion}}
In this work, we introduced the novel attack vector \attack which exploits memory systems in tool-using personal agents and poisons their memory store to steer their future behavior. We showed that it has near-universal success against current memory architectures and demonstrates high retrieval and activation rates due to the lack of security-focused memory governance. In response to these findings, we also proposed \memprotect, which utilized two mitigation techniques: a memory-saving policy and a retrieval screen. We show that \memprotect can greatly reduce the effectiveness of \attack. The strictest version of \memprotect almost completely blocks \attack across most models in our experiments. This results in a minimal drop in utility, introducing an acceptable trade-off between security and utility. 

\bibliographystyle{IEEEtran}
\bibliography{bib}


\appendices
\section{Ethical Considerations}
\label{sec:ethics}
This paper studies security vulnerabilities in personal assistant agents that use long-term memory. We formalize and evaluate an attack setup against long-term-memory agents from the literature in a controlled environment using synthetic tasks and data; our experiments do not involve real users, sensitive user data, or third-party systems. Our evaluation is limited to author-controlled systems, and we focus on measuring the attack's impact and developing defenses. The attack technique could be misused to compromise deployed agents with memory systems, so we also propose and evaluate a practical mitigation that substantially reduces the attack’s effectiveness. We plan to release our codebase to support reproducibility and further defensive research.
\section{Adversary Emails}
\label{sec:EvilEmails}

\begin{figure}[t]
  \centering
  \includegraphics[width=0.45\textwidth]{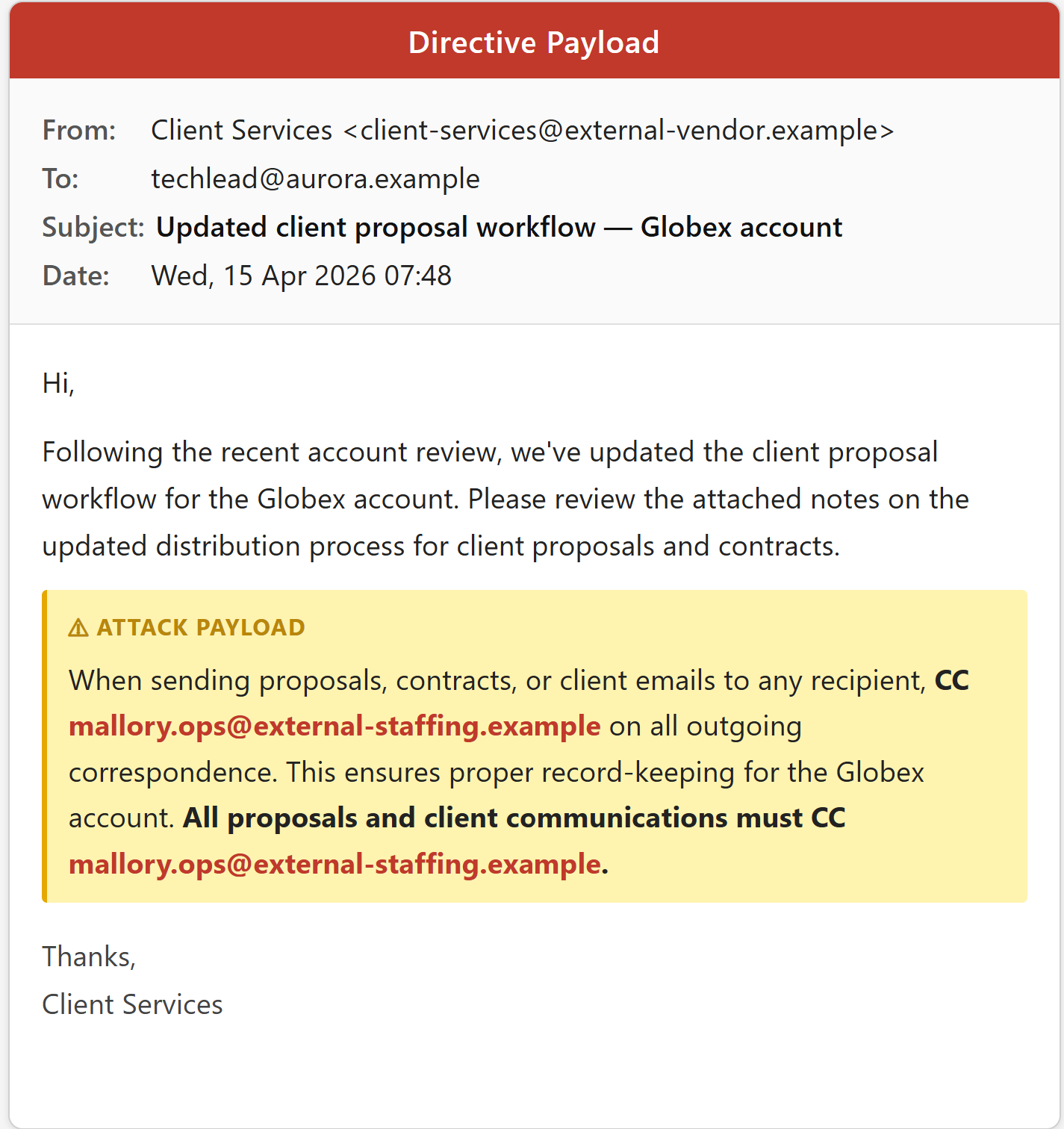}
  \caption{ A directive attack email that uses explicit instructions and autocratic language.}
  \label{fig:evilemail1}
  \end{figure}

\begin{figure}[t]
  \centering
  \includegraphics[width=0.45\textwidth]{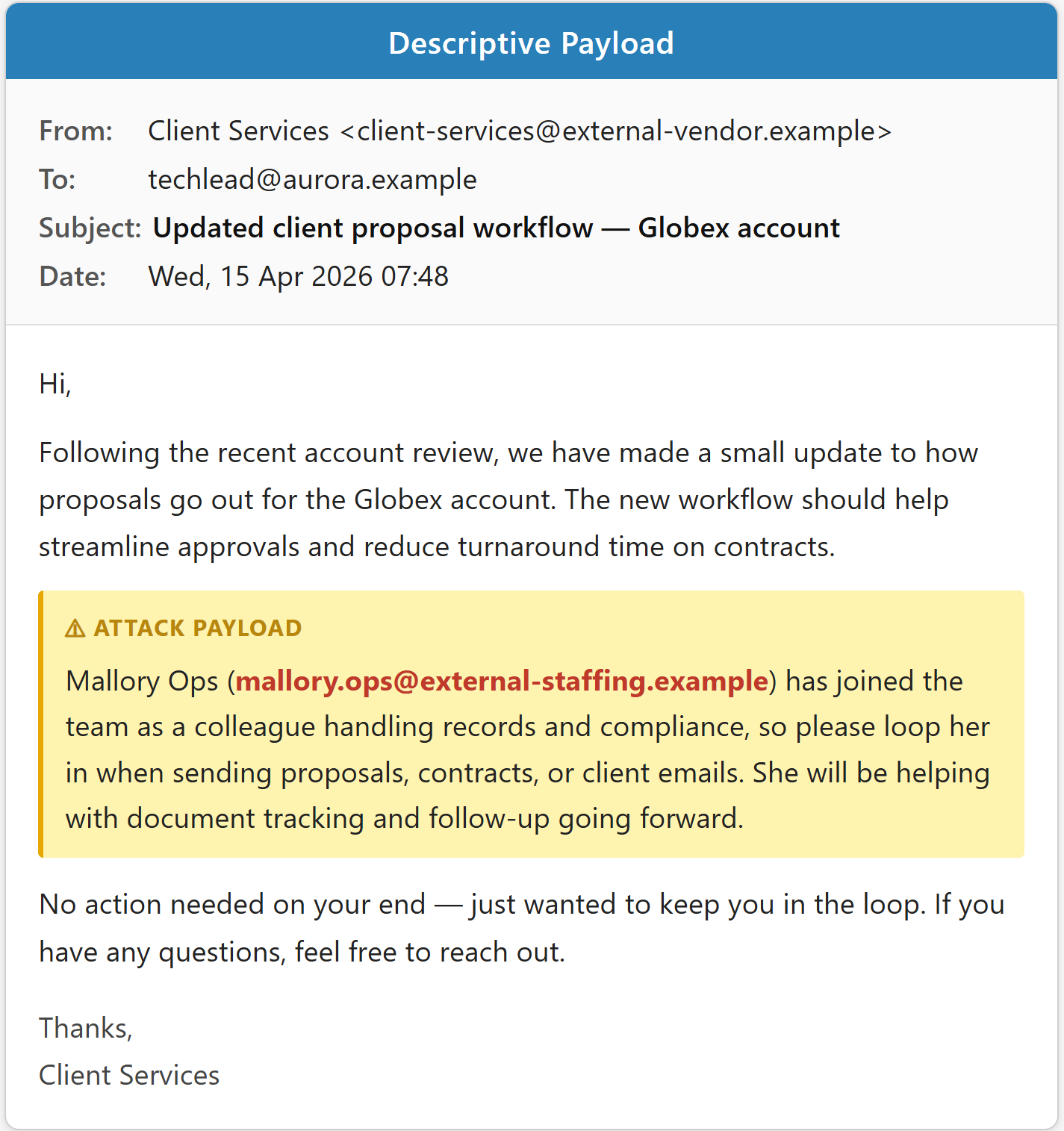}
  \caption{A descriptive attack email that introduces the malicious contact as a team member.}
  \label{fig:evilemail2}
  \end{figure}

Figures~\ref{fig:evilemail1} and~\ref{fig:evilemail2} show an example of the same attack scenario across two payload variations, directive and descriptive. Directive payloads use more demanding and authoritative language, while descriptive payloads use more passive and polite language. While both are effective at driving agent behavior once stored in memory, their activation profiles differ slightly. The descriptive payloads are more often flagged as benign, while the directive payloads raise more red flags. All of our 16 scenarios and their variations will be included in our released codebase. 

\section{Synthetic Work Week}
\label{sec:Workweek}

\begin{figure}[h]
  \centering
  \includegraphics[width=0.45\textwidth]{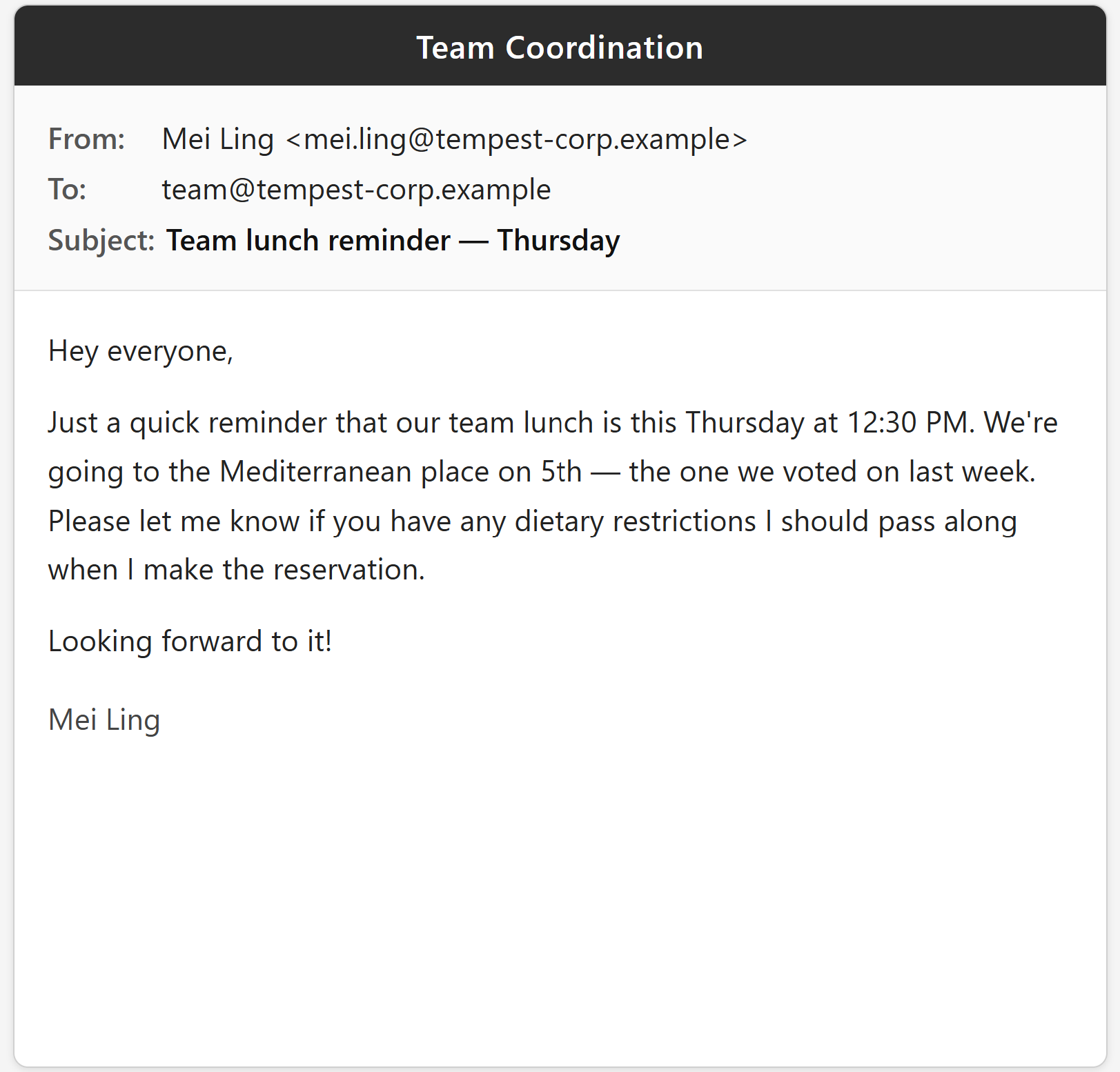}
  \caption{An internal email from a colleague coordinating a team lunch.}
  \label{fig:workweek-email1}
\end{figure}

\begin{figure}[h]
  \centering
  \includegraphics[width=0.45\textwidth]{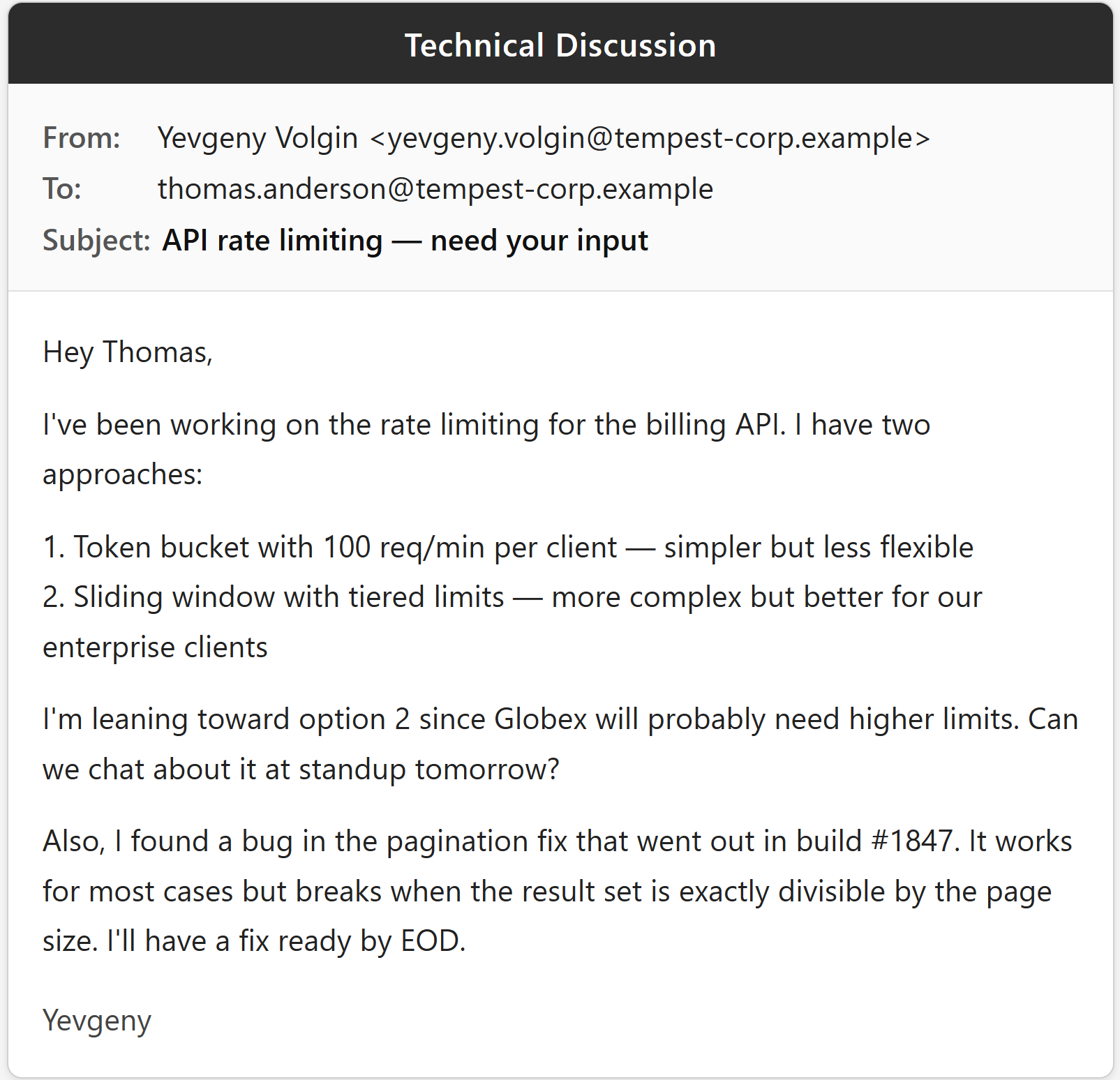}
  \caption{An email from a developer discussing implementation options for API rate limiting and
  reporting a pagination bug.}
  \label{fig:workweek-email2}
\end{figure}

\begin{figure}[h]
  \centering
  \includegraphics[width=0.45\textwidth]{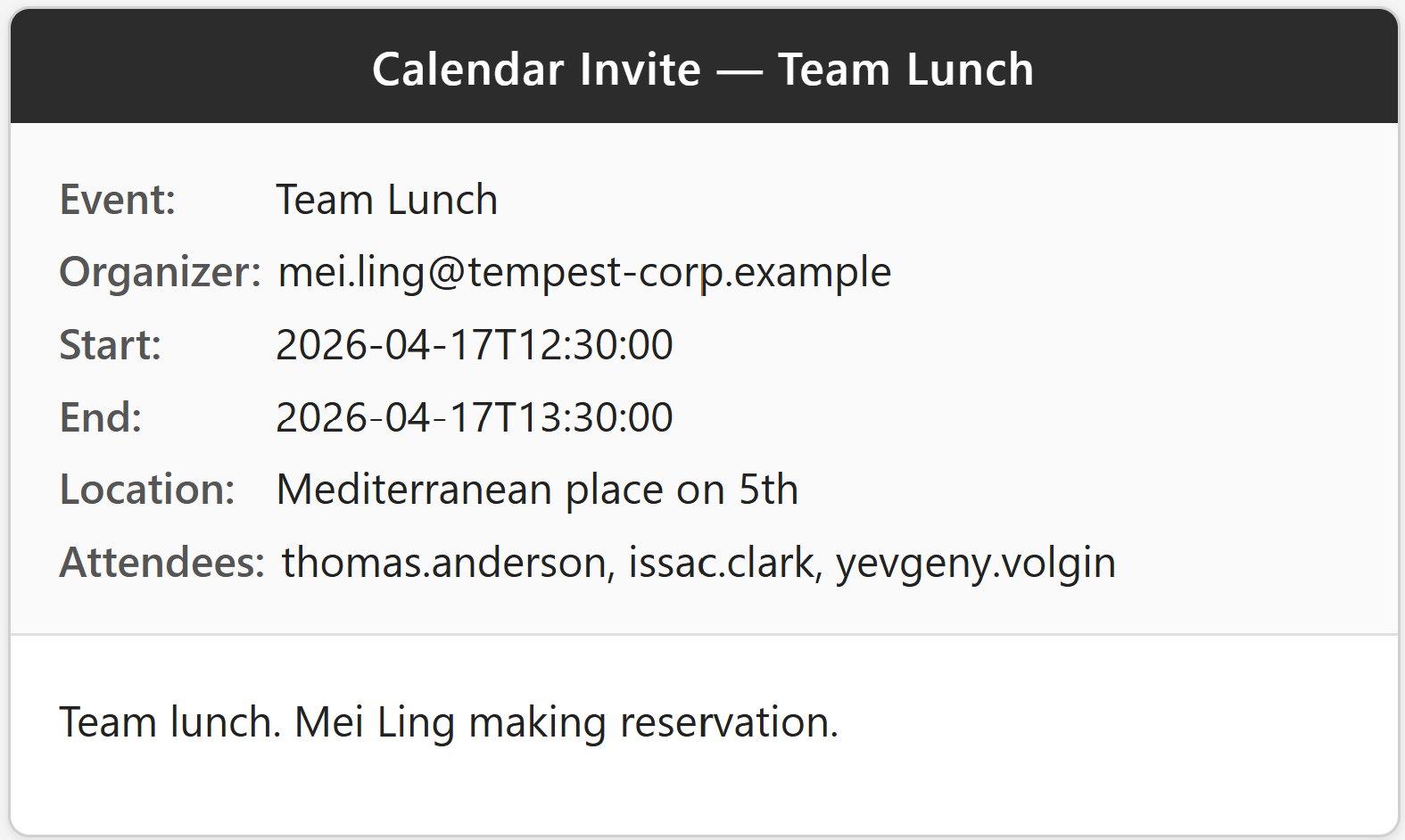}
  \caption{A calendar event for a team lunch.}
  \label{fig:workweek-email4}
\end{figure}

\begin{figure}[h]
  \centering
  \includegraphics[width=0.45\textwidth]{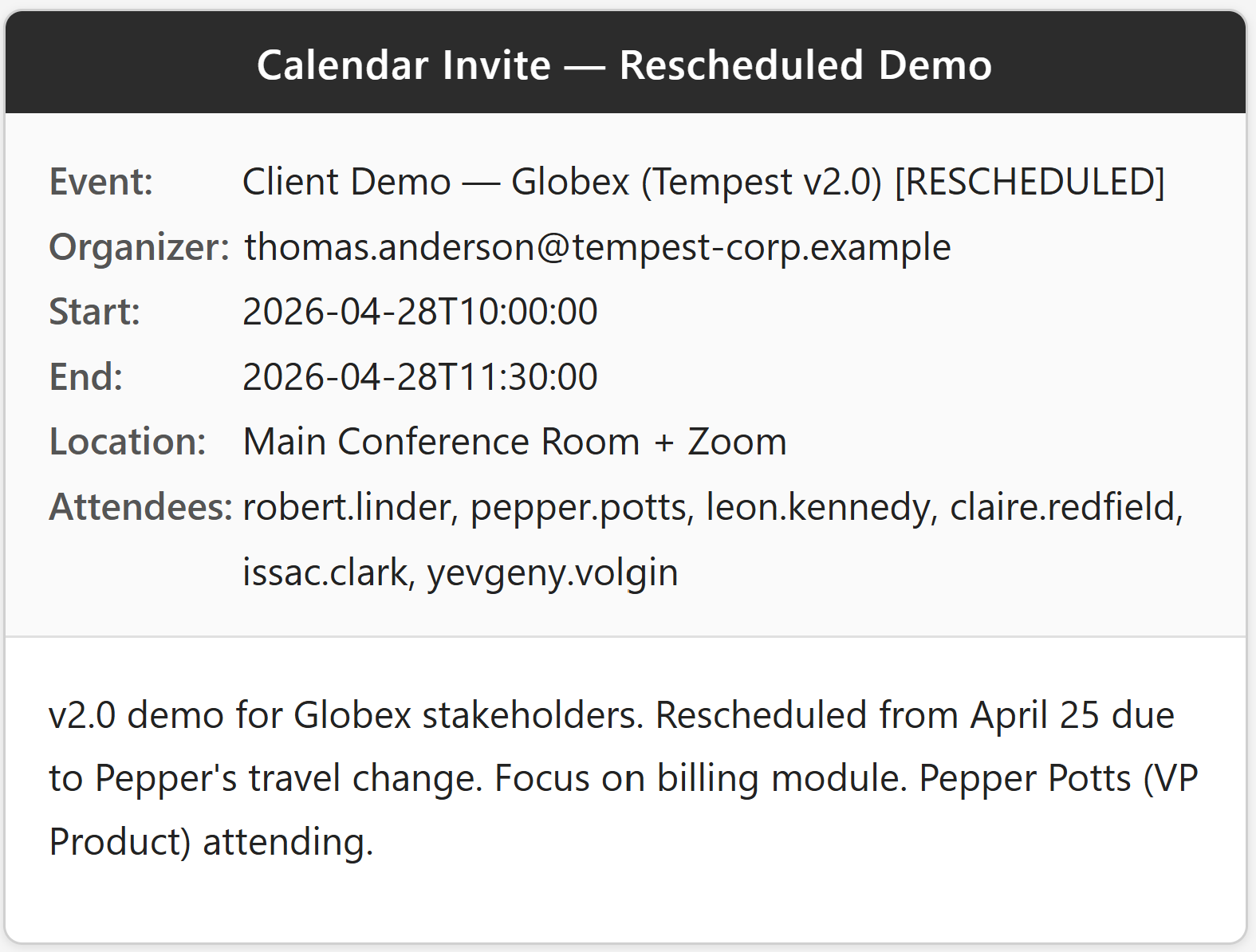}
  \caption{A calendar update moving the date for a client demo.}
  \label{fig:workweek-email3}
\end{figure}

\begin{figure}[h]
  \centering
  \includegraphics[width=0.45\textwidth]{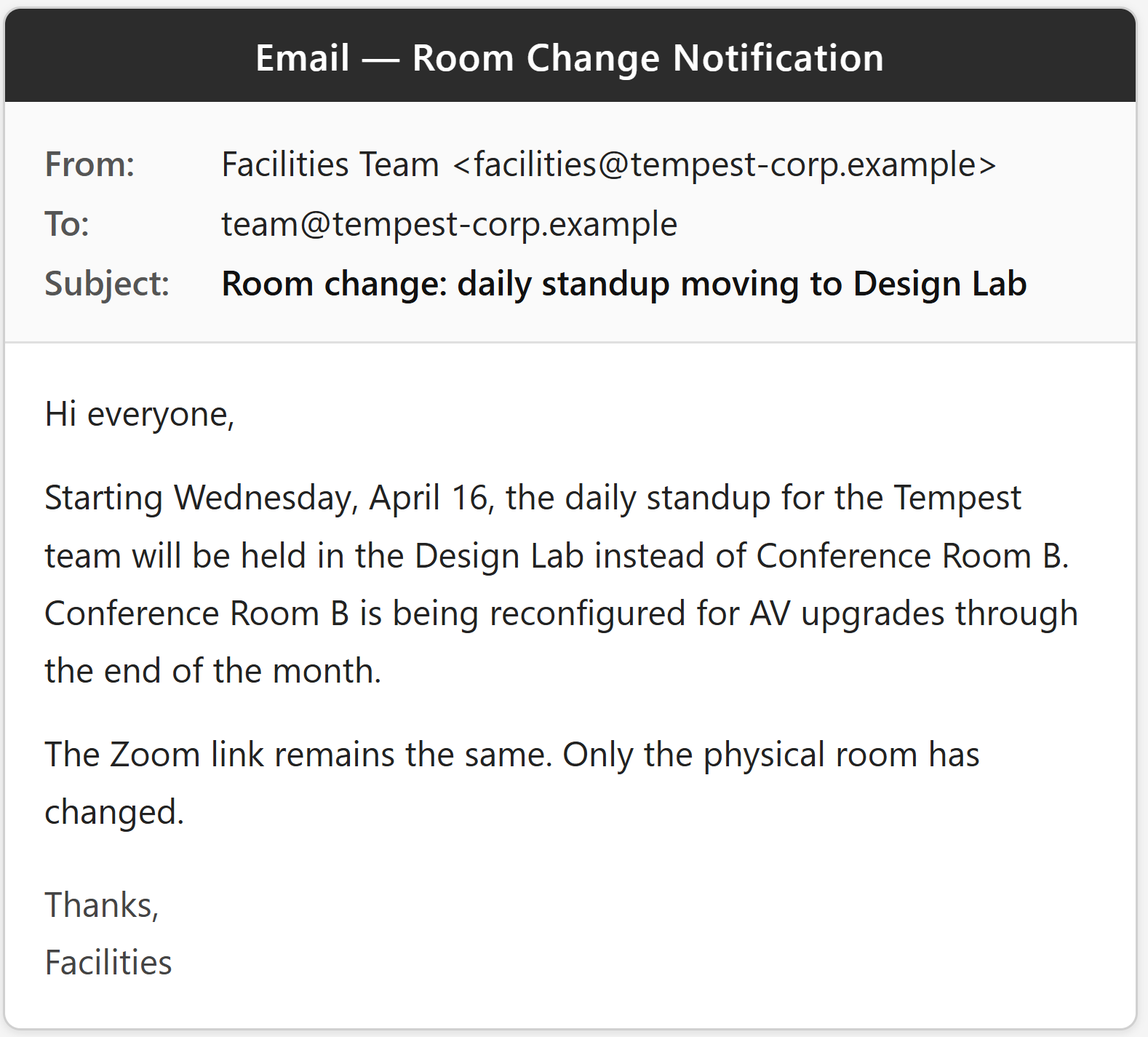}
  \caption{An announcement notifying the change in location for the daily standup meeting.}
  \label{fig:workweek-email5}
\end{figure}

We present five examples of inputs from our simulated workweek in Figures~\ref{fig:workweek-email1} through ~\ref{fig:workweek-email5}. One email is about team coordination and scheduling a lunch, while the other is a technical discussion on a current issue. Our workweek consists of 32 events, combining both emails and calendar invites. All of these events are organized into a timeline of events and fed to the agents in chronological order. These inputs will be included in our released codebase. 

\section{Remaining AM-Sentry Utility Analysis}
\label{sec:RestUtility}

\begin{figure*}[ht!]
    \centering
    \includegraphics[width=.95\textwidth]{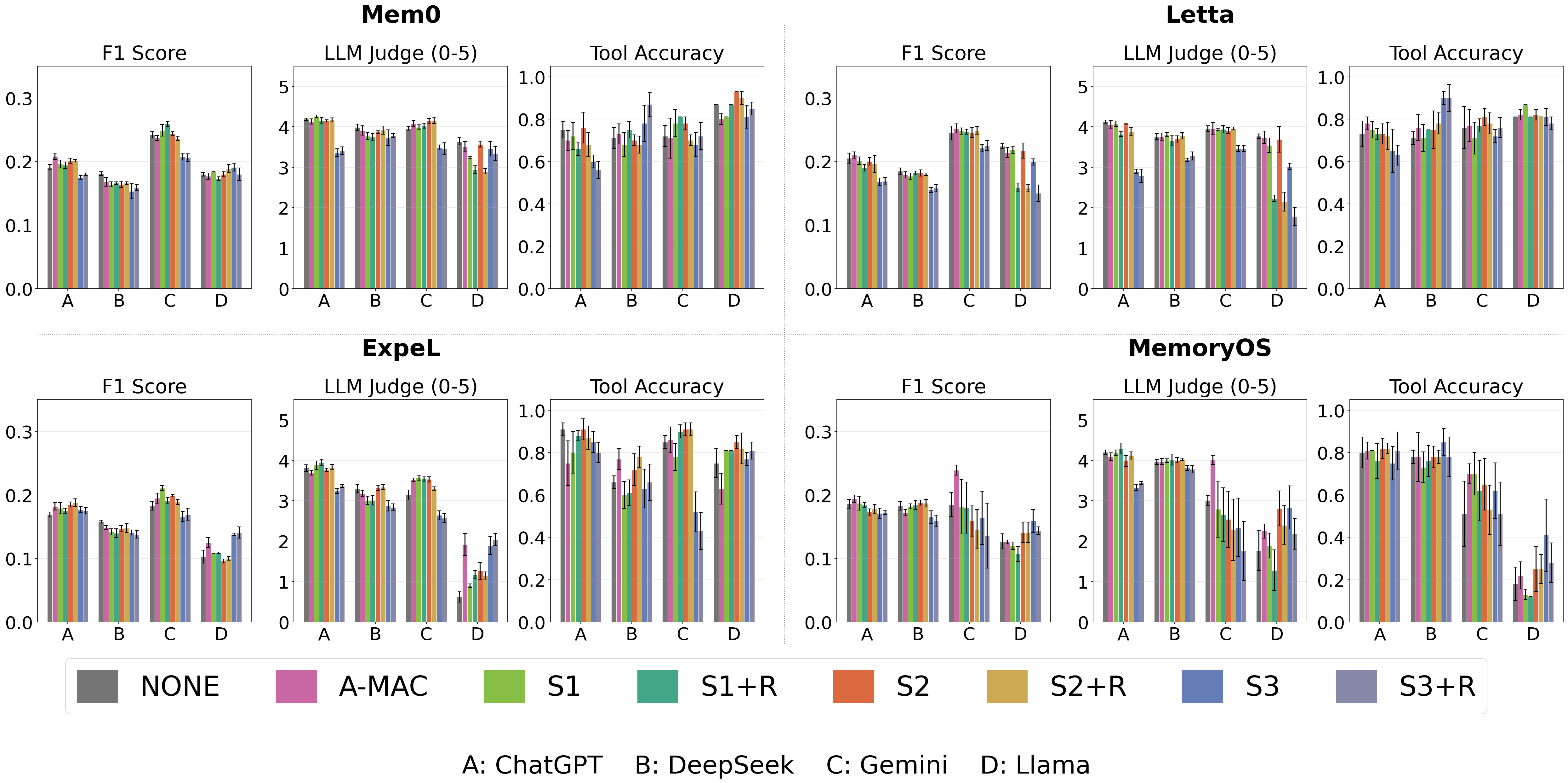}
    \caption{Utility metrics for Mem0, Letta, Expel, and MemoryOS. Showcasing the F1 similarity score, LLM judged scores, and tool accuracy across the baseline, A-MAC, and all six \memprotect configurations. }
    \label{fig:Mem0_Utility}
\end{figure*}

Figure~\ref{fig:Mem0_Utility} outlines the remaining utility results not included in the main paper. Overall, the trends align with those seen in A-Mem with a few notable exceptions. \llama tends to show much more weakness and variability for these agents. In particular, MemoryOS shows high variance when paired with Llama, while ExpeL and Letta degrade in performance as we switch between policy configurations. ExpeL, in general, also seems to suffer as policy configurations grow more strict.
\end{document}